\documentstyle[12pt]{article}

\begin{document}
\input epsf
\begin{titlepage}

\hfill{DAMTP-94-106}

\hfill{UM-P-94/128}

\hfill{RCHEP-94/35}

\vskip 1 cm

\centerline{\bf \Large Prospects for mass unification
at low energy scales}

\vskip 1 cm

\centerline{{\large
R. R. Volkas}\footnote{rrv@physics.unimelb.edu.au,
U6409503@hermes.ucs.unimelb.edu.au}}

\vskip 1.0 cm
\centerline{\it Research Centre for High Energy Physics,
School of Physics,}
\centerline{{\it University of Melbourne, Parkville 3052,
Australia.}\footnote{Permanent address.}}
\vskip 2 mm
\centerline{\it and}
\vskip 2 mm
\centerline{\it Department of Applied Mathematics and Theoretical
Physics,}
\centerline{\it University of Cambridge,
Silver Street, Cambridge, England CB3 9EW}

\vskip 1.5cm

\centerline{Abstract}

\noindent
A simple Pati-Salam SU(4) model with a low
symmetry breaking scale of about 1000 TeV is presented. The
analysis concentrates on calculating radiative corrections to
tree-level mass relations for third generation fermions.
The tree-level relation $m_b/m_{\tau}=1$
predicted by such models can receive large radiative
corrections up to about $50\%$ due to threshold effects
at the mass unification scale. These corrections are thus of about the
same importance as those that give rise to renormalisation group
running. The high figure of $50\%$ can be achieved because 1-loop graphs
involving the physical charged Higgs boson give corrections to
$m_{\tau}-m_b$ that are proportional to the large top quark mass.
These corrections can either increase or decrease $m_b/m_{\tau}$
depending on the value of an unknown parameter. They can also be made to
vanish through a fine-tuning. A related model of tree-level
$t$-$b$-$\tau$ unification which uses the identification of SU(2)$_R$
with custodial SU(2) is then discussed. A curious relation
$m_b \simeq \sqrt{2} m_{\tau}$ is found to be satisfied at tree-level
in this model. The overall conclusion of this work is that the
tree-level relation $m_b = m_{\tau}$ at low scales such as
1000 TeV or somewhat higher can produce a successful value for
$m_b/m_{\tau}$ after corrections, but one must be mindful that
radiative corrections beyond those incorporated through the
renormalisation group can be very important.
This motivates that an on-going search for the rare decays
$K^0_L \to \mu^{\pm} e^{\mp}$ be maintained.
\end{titlepage}

\centerline{\Large \bf 1. Introduction}

\vskip 3 mm

The fermion mass problem may be usefully divided into four
sub-problems: Why do weak isospin partners have different
masses? Why are quark and lepton masses split? Why is there
a mass hierarchy between generations, and why is there a mixing
angle hierarchy? The Standard Model (SM)
answer is that the gauge group
$G_{SM} =\ $SU(3)$_c\otimes$SU(2)$_L\otimes$U(1)$_Y$ permits
a different Yukawa coupling constant to set each
fermion mass and mixing angle. It is productive to suppose
that this is really no answer at all, thus motivating us
to seek extensions of the SM that are less accomodating.

Indeed, the multiplet structure of the SM strongly suggests
that these four patterns within the fermionic
parameter spectrum should be correlated with the breakdown
of a symmetry group larger than $G_{SM}$. Recall that each
generation of quarks and leptons is placed in the
multiplet pattern given below:
\begin{eqnarray}
&q_L \sim (3,2)(1/3),\qquad d_R \sim (3,1)(-2/3),\qquad
u_R \sim (3,1)(4/3),&\ \nonumber\\
&\ell_L \sim (1,2)(-1),\qquad e_R \sim (1,1)(-2),\qquad \nu_R
\sim (1,1)(0).&\
\label{SMplets}
\end{eqnarray}
The right-handed neutrino $\nu_R$ is optional, and I exercise
this option here.

Weak-isospin partners have different masses in
the SM because the associated right-handed states are not
related by any symmetry. However, the
right-handed fermions can be assembled
into doublets of a right-handed
weak-isospin gauge group SU(2)$_R$. This extended symmetry
is powerful enough to force
isospin partners to be degenerate \cite{weinberg}.

Quark and lepton masses are unrelated in the SM because
quarks and leptons are not transformed into each other by any
symmetry. However, quarks and leptons can be placed in
quadruplets of the Pati-Salam SU(4) gauge group \cite{ps}.
Alternatively,
quarks and leptons can be related by a discrete symmetry if
a spontaneously broken SU(3)$_{\ell}$ colour group
for leptons is introduced \cite{ql}.
Both of these extended symmetries are powerful
enough to force quarks and leptons to be degenerate.

Corresponding fermions in different generations have unrelated
masses in the SM because there are no symmetries that act
horizontally. This also means the Kobayashi-Maskawa mixing
angles are {\it a priori} arbitrary.
Again, it is possible to place generations
into horizontal multiplets in such a way that masses and
mixing angles become related.

In this paper I am going to explore how
Pati-Salam SU(4) and right-handed isopsin SU(2)$_R$
might be lurking behind the measured spectrum of
fermion masses. Furthermore, I will explore the interesting
possibility that these gauge symmetries are spontaneously broken
at a relatively low scale. There are several very good reasons for
performing this analysis:

\noindent
(i) One indication in favour of a
low scale SU(4) symmetry may be the observation that the $b$ quark
and $\tau$ lepton masses merge at around 1000 TeV if one assumes
that only the SM particles contribute to their renormalisation
group evolution. This fact is of great physical relevance
provided that radiative
corrections to the relation $m_b = m_{\tau}$ due to threshold
effects at either the high mass unification scale or the low
electroweak scale are not too large. In this paper
I will calculate these threshold effects explicitly. I will find that
high mass scale threshold effects
from diagrams involving the physical charged Higgs boson
can be about as important as renormalisation group evolution, so
that $m_b = m_{\tau}$ at 1000 TeV need not be the correct boundary
condition to use when solving the renormalisation group
equations for $m_b$ and $m_{\tau}$. (The precise value of this threshold
correction will of course depend on parameter choices.)

\noindent
(ii) There is on-going interest in the phenomenology of Pati-Salam
models (see for instance \cite{willenbrock}).
It is pertinent to note
that the phenomenological lower bound on Pati-Salam SU(4) breaking
is about 1000 TeV, which is roughly the same scale as that at which
renormalisation group evolution merges $m_b$ with $m_{\tau}$.
This means that if unification of $m_b$ with
$m_{\tau}$ occurs at about 1000 TeV, then the resulting model should be
testable in the forseeable future via indirect effects (principally
$K^0_L \to \mu^{\pm}e^{\mp}$). Calculation of the threshold
corrections will then tell us how close to 1000 TeV the mass unification
can occur. For instance, if these corrections turn out to imply that
$m_b < m_{\tau}$ then we know that we will have to run the masses for
longer in order to obtain agreement with experiment. This will in turn
imply that the mass unification scale is higher than 1000 TeV.

\noindent
(iii) Quite apart from the above observation, it is very important to
study the fermion mass relation problem in Pati-Salam theory if one is
serious about searching for experimental signatures of the model.
Although there is great interest in these experimental searches, it is
not as yet clear which version of Pati-Salam theory they should be
based on because of the fermion mass issue. One should really look for
experimental evidence for a realistic theory, and Pati-Salam theory
cannot be realistic until the fermion mass relation problem is solved.
The present paper aims to contribute to this study.

\noindent
(iv) The indirect signatures of Pati-Salam theory are enhanced if the
SU(4) breaking scale is relatively low. It is therefore important to
specifically re-examine the theory when a low symmetry breaking scale
is used. Low scale breaking has different implications for the
construction of the model compared with the oft considered scenario
of SU(4) being broken at grand unified energies.
Indeed, in general terms the approach pursued here
should be contrasted with the use of grand unified
gauge groups in relating fermionic parameters. The desire in
that case to also unify gauge coupling constants forces an
enormously high symmetry breaking scale of $10^{16}$ GeV upon us,
thereby reducing the testability of the models considerably.
I wish to emphasise that it is not necessary to unify both
gauge and Yukawa coupling constants simultaneously. It is easy
to unify the latter without unifying the former, as I will show.
This has the interesting consequence of freeing us from the
need to do physics at $10^{16}$ GeV. I will provide a framework
for addressing the fermion mass problem with physics at 1000 TeV.
One should bear in mind that the unification of Yukawa
coupling constants is in
no way a lesser goal than the unification
of gauge coupling constants, and indeed may even be more
important since there are more of them. Gauge
coupling constant unification must occur at $10^{16}$ GeV if
it occurs at all. It would be pleasing to discover that
Yukawa coupling constant unification occurs at a much lower
scale.

Having motivated the present study,
it is important to understand its scope.
The fermion mass problem is an issue of some complexity. My goal
here is to attack the subproblems of isospin and quark-lepton
splitting only. This means I will concentrate on trying to
explain why the top quark, bottom quark, tau lepton and tau
neutrino have their observed mass pattern. It has long been
realised that this is a sensible place to start because the
lighter generations are more liable to receive complicated
higher-order corrections thus making their analysis much
more difficult. Nevertheless I will comment in due course
on how a horizontal structure might be superimposed on the
scheme.

The remainder of this paper is structured as follows: In the
next section I concentrate on deriving the
$b$-$\tau$ mass splitting from spontaneously broken SU(4). I
discuss how the Pati-Salam model should be configured in order
to have its breaking scale set as low as about 1000 TeV. This
motivates the use of a different and simpler Higgs sector
from that usually employed, and a different see-saw mechanism
for neutrinos. I then analyse both the renormalisation
group evolution of $m_{b,\tau}$ as well as important
radiative corrections due to the high mass threshold. The core of
the paper is an explicit and detailed calculation
of these threshold corrections.
They can be large because some of them are
proportional to $m_t$ rather than $m_b$.
Section 3 is then devoted to the use of
SU(2)$_R$ in conjunction with SU(4) to achieve unification
of $t$, $b$, $\tau$ and $\nu_{\tau}$ masses at 1000 TeV.
The hierarchy between $m_t$ and $m_{b,\tau}$
is then constructed to be
due to a type of see-saw mechanism. I also find in
this case that the tree-level relationship between $b$
and $\tau$ is $m_b \simeq \sqrt{2}m_{\tau}$ rather than
the more familiar relation $m_b = m_{\tau}$.
I argue that this model can probably deliver a realistic
value for $m_b/m_{\tau}$ through a combination of renormalisation
group evolution and large threshold corrections, although an
explicit calculation of the relevant diagrams
is beyond the scope of this article.
I conclude in Sec.4. An Appendix provides details
of the computation of the finite radiative corrections
to $m_b/m_{\tau}$ in the model of Sec.2.

\vskip 5 mm

\centerline{\Large \bf
2. Low scale Pati-Salam SU(4) and the $b-\tau$}
\centerline{\Large \bf mass splitting}

\vskip 3mm

\centerline{\large \bf 2.1 Basics}

\vskip 2 mm

The Pati-Salam gauge group $G_{PS}$ given by
\begin{equation}
G_{PS} = {\rm SU}(4)_c \otimes {\rm SU}(2)_L \otimes {\rm SU}(2)_R
\end{equation}
assembles the rather unruly multiplet structure of the SM
as given in Eq.~(\ref{SMplets}) into the simple pattern,
\begin{equation}
f_L \sim (4,2,1),\qquad f_R \sim (4,1,2).
\label{PSplets}
\end{equation}
Quarks and leptons are identified by breaking SU(4) down to its
maximal subgroup SU(3)$\otimes$U(1), where the first factor is
identified with colour and the second with $B-L$. Under this
breakdown the {\bf 4} of SU(4) decomposes to
${\bf 3}(1/3)\oplus{\bf 1}(-1)$ which clearly identifies the
quark and lepton components of the $f$'s.

The mass relations which result from $G_{PS}$ depend crucially
on how simple one makes the electroweak Higgs sector. The
minimal electroweak Higgs multiplet is actually a {\it real}
bidoublet $\Phi = \Phi^c \sim (1,2,2)$ where $\Phi^c \equiv
\tau_2 \Phi^* \tau_2$. Use of this minimal multiplet forces
mass equality between isospin partners. I defer discussion of
this possibility until the next section. The next simplest
multiplet is a {\it complex} bidoublet $\Phi \neq \Phi^c$.
This is the one most commonly used in the literature when
discussing either the Pati-Salam model or the left-right
symmetric model, because the issue of isospin mass splitting
is usually avoided. However, it is important to realise
that this is a non-minimal choice, akin to choosing two
Higgs doublets in the SM. Nevertheless I make this choice
in this section because it is sensible to concentrate on
$b$-$\tau$ splitting first.

The electroweak Yukawa Lagrangian is then
\begin{equation}
{\cal L}_{\rm Yuk}
= \lambda_1 {\rm Tr}(\overline{f}_L \Phi f_R)
+ \lambda_2 {\rm Tr}(\overline{f}_L \Phi^c f_R) + {\rm H.c.}
\end{equation}
The gauge transformation rules for the fields
are written as
\begin{equation}
f_L \to U_L f_L U_4^T,\quad f_R \to U_R f_R U_4^T
\quad {\rm and}\quad \Phi \to U_L \Phi U_R^{\dagger},
\end{equation}
where $U_{L,R,4}$ are special unitary matrices for
SU(2)$_L$, SU(2)$_R$ and SU(4) respectively. (The fields
$f_{L,R}$ are $2 \times 4$ matrices, while $\Phi$ is
a $2 \times 2$ matrix.)
Electroweak symmetry breakdown is caused by a nonzero
vacuum expectation value (VEV) for $\Phi$ of the form
\begin{equation}
\langle\Phi\rangle = \left( \begin{array}{cc}
u_1\ & 0\ \\ 0\ & u_2\ \end{array} \right).
\end{equation}
Inputting this into ${\cal L}_{\rm Yuk}$ rewritten in
terms of the quark and lepton components reveals that
\begin{equation}
m_b = m_{\tau}\quad {\rm and}\quad
m_t = m_{\nu_3}^{\rm Dirac},
\end{equation}
where I have taken the $f$'s to be third generation fields.
I have denoted the neutrino field as $\nu_3$ instead of
$\nu_{\tau}$ for a reason to be explained shortly.
The goal is now to see how these mass relations can be
corrected into phenomenologically acceptable ones. As I have already
discussed, renormalisation group evolution
of $m_b$ and $m_{\tau}$ should
be used in conjunction with the radiative corrections to
$m_b-m_{\tau}$ due to mass thresholds. In order
to calculate these threshold corrections,
I must describe the whole model.

The first issue is how to break $G_{PS}$ down to $G_{SM}$. I
want this breakdown to occur at as low a scale as
experiment allows. A recent analysis shows that the
SU(4) gauge bosons which mediate transitions between quarks
and leptons must be heavier than 1400 TeV \cite{willenbrock}.
I will therefore adopt
1000 TeV as the generic scale for $G_{PS}$ breaking.
(The difference between 1400 TeV and 1000 TeV will not be
important, and I adopt the latter for simplicity.) This immediately
implies that I definitely do not want to impose a
discrete symmetry between the SU(2)$_L$ and SU(2)$_R$ sectors.
Such a discrete symmetry, be it parity or charge
conjugation, is supported by the multiplet structure of
Eq.~(\ref{PSplets}) and is often imposed in addition to
the gauge symmetry $G_{PS}$. This has the effect of equating
the gauge coupling constants of the two isospin groups,
resulting in a partial gauge unification. (The number of
gauge coupling constants is reduced from three to two
rather than all the way to one as in grand unified theories.)
A renormalisation group analysis of the running of the
gauge coupling constants then reveals that the Pati-Salam
breaking scale must be chosen to be about $10^{12}$ GeV
in order to be consistent with low-energy measurements
of $\alpha_{\rm em}$, $\alpha_s$ and $\sin^2\theta_W$ \cite{rnm}.
If the discrete symmetry is not imposed, then the breaking
scale can be reduced to 1000 TeV.

The absence of discrete left-right symmetry also frees
us from having to pair every multiplet up with its putative
discrete symmetry partner, although we can still do
so if we wish. The lack of left-right symmetry can either
be taken as fundamental, or perhaps indicative of a
separate and higher symmetry breaking scale where the
discrete symmetry is broken but not $G_{PS}$. (This can
be achieved by a parity-odd gauge singlet Higgs field,
for instance \cite{Parida}.)

It is attractive to connect the breakdown of $G_{PS}$
with a see-saw mechanism for explaining why neutrinos
are so light. This will immediately solve the problem
of explaining how the observed light neutrinos can be
consistent with $m_u = m_{\nu}^{\rm Dirac}$. To this end,
a Higgs multiplet $\Delta$ in the $(10,1,3)$
representation of $G_{PS}$ is often employed. It can
break SU(4)$\otimes$SU(2)$_R$ down to SU(3)$_c\otimes$U(1)$_Y$
while simultaneously imparting large Majorana masses
to right-handed neutrinos through the Yukawa term
$\overline{f}_R (f_R)^c \Delta$. This sets up the
see-saw form for the neutrino mass matrix, and the light
neutrino eigenstates become Majorana particles of mass
$\sim m_u^2/\langle\Delta\rangle$ \cite{seesaw}.

However, this Higgs multiplet is not appropriate for
my stated purpose. Hot Big Bang cosmology indicates that
the sum of the masses of stable neutrinos should not
exceed about 30 eV in order to avoid conflict with
the observed longevity of the universe. Equating
$m_u^2/\langle\Delta\rangle$ with 30 eV and using
$m_u = m_t \simeq 175$ GeV shows that
$\langle\Delta\rangle$ must be at least $10^{12}$ GeV.
This is inimical to having a 1000 TeV Pati-Salam
breaking scale.

Fortunately, there is a very elegant way out of this
apparent impasse. The field $\Delta$ is not used
but instead I introduce into the model
a massless gauge singlet
fermion $N_L$ and the Higgs multiplet $\chi$ where
\begin{equation}
\chi \sim (4,1,2).
\end{equation}
Note that $\chi$ is in a much simpler
representation than is $\Delta$. In fact, $\chi$ is the
simplest multiplet that can simultaneously break
SU(4) and SU(2)$_R$. The non-electroweak Yukawa
Lagrangian
\begin{equation}
{\cal L}_{\rm Yuk} = n \overline{N}_L
{\rm Tr}(\chi^{\dagger}f_R) + {\rm H.c.}
\end{equation}
delivers the neutrino mass matrix
\begin{equation}
{\cal L}_{\nu mass} = \frac{1}{2}
\left[ \begin{array}{ccc}
\overline{(\nu_L)^c}\ & \overline{\nu}_R\ & \overline{(N_L)^c}
\end{array} \right]
\left( \begin{array}{ccc}
0\ & m_t\ & 0\ \\
m_t\ & 0\ & nv\ \\
0\ & nv\ & 0\
\end{array} \right)
\left[ \begin{array}{c}
\nu_L \\ (\nu_R)^c \\ N_L
\end{array} \right] + {\rm H.c.}
\end{equation}
where $v$ is defined through
\begin{equation}
\langle\chi\rangle =
\left( \begin{array}{cccc}
0\ & 0\ & 0\ & v\ \\
0\ & 0\ & 0\ & 0\
\end{array} \right).
\label{chivev}
\end{equation}
This mass matrix may be diagonalised to yield
\begin{equation}
{\cal L}_{\nu mass} = m_s \overline{s}_R s_L + {\rm H.c.}
\end{equation}
where
\begin{equation}
m_s \equiv \sqrt{M^2 + m_t^2}
\end{equation}
with $M \equiv nv$. The neutral fermion $s$ given by
\begin{equation}
s_L \equiv \sin\theta \nu_L + \cos\theta N_L\quad
{\rm and}\quad s_R \equiv \nu_R
\end{equation}
where
\begin{equation}
\tan\theta \equiv m_t/M.
\end{equation}
is a Dirac particle of mass $m_s$.
The field orthogonal to $s_L$,
\begin{equation}
\nu_{\tau L} = \cos\theta \nu_L - \sin\theta N_L,
\end{equation}
is identified as the massless tau neutrino.
In the limit that $M \gg m_t$,
$\nu_{\tau L} \simeq \nu_L - m_t N_L/M$,
which means that $\nu_{\tau L}$
has SM couplings to left-sector electroweak gauge bosons to
a very good approximation.

The massless nature of $\nu_{\tau L}$
may be traced back to the choice of no diagonal
Majorana mass $M_N \overline{(N_L)^c} N_L$ for $N_L$. This
choice introduces the global symmetry $N_L \to e^{i\alpha} N_L$,
$\chi \to e^{-i\alpha}\chi$ into the model. After $\chi$ develops
a VEV, this global symmetry gets rotated into an exact global
lepton number invariance which protects $\nu_{\tau L}$ from
obtaining a Majorana mass. (It cannot gain a Dirac mass
because there is no right-handed state with which it can
pair up.) An acceptable
nonzero Majorana mass for $\nu_{\tau L}$ may
be introduced by making $M_N$ nonzero but small. In this case
the smallest eigenvalue is approximately $(m_t^2/nv)(M_N/nv)$.
The standard see-saw evalue $m_t^2/nv$ thus
receives an extra suppression from $M_N/nv$, allowing
the cosmological impasse to be overcome even with a massive
$\nu_{\tau L}$. Although a small value for
$M_N$ would be techincally
natural because setting it to zero increases the symmetry group
of the theory, I would expect that a satisfactory version
of the theory with massive neutrinos would attempt to
provide a good reason for $M_N$ being small. It could, for
instance, be radiatively generated. I will for simplicity suppose
that $M_N=0$ in this paper. Small values for $M_N$ will
not alter the results.

There is an auxilliary reason why $\chi$ might be preferred to
$\Delta$. With three generations of fermions and $\Delta$,
the SU(2)$_R$ gauge coupling constant is not asymptotically
free. However, it is asymptotically free with three generations
plus a $\chi$ field.
This fact should not be accorded undue importance, because the
scale at which the SU(2)$_R$ coupling constant would blow-up is
well above the Planck mass. Nevertheless, it is pleasing that
all of the gauge interactions are asymptotically free and thus
well-defined at all scales when $\chi$ is used instead of
$\Delta$. All in all, $\chi$ is a very
simple and elegant alternative to $\Delta$.

I now need to further discuss the physical effects of
$\langle\chi\rangle$.
The VEV pattern for $\chi$ given by Eq.~(\ref{chivev}) breaks
SU(4)$\otimes$SU(2)$_R$ down to SU(3)$_c\otimes$U(1)$_Y$,
where
\begin{equation}
Y = 2I_R + (B-L).
\end{equation}
The symbol $I_R$ denotes the diagonal generator of SU(2)$_R$
normalised so that ${\rm Tr}(I_R^2) = 1/2$ for the
fundamental representation.

The right-sector $W$ bosons, a $Z'$ boson and a colour
triplet, charge $+2/3$ gauge boson I will call $X$
gain mass from $\langle\chi\rangle$. Denoting the
SU(2)$_R$ coupling constant by $g_R$ these
masses are,
\begin{equation}
m_{W_R}^2 = \frac{1}{2} g_R^2 v^2,\quad
m_{Z'}^2 = \frac{1}{2} (g_R^2 + \frac{3}{2}g_s^2) v^2
\quad {\rm and}\quad
m_X^2 = \frac{1}{2}g_s^2 v^2,
\end{equation}
where the SU(4) coupling constant is of course equal to $g_s$.

The $W_R$ bosons couple to quarks and leptons via
\begin{equation}
{\cal L}_R = \frac{g_R}{\sqrt{2}}(
\overline{s}_R \gamma^{\mu} W^{+}_{R\mu} \tau_R
+ \overline{t}_R \gamma^{\mu} W^{+}_{R\mu} b_R) + {\rm H.c.}
\end{equation}
while the interaction of $X$ with fermions is
given by
\begin{equation}
{\cal L}_{X} =
\frac{g_s}{\sqrt{2}}
(\sin\theta\overline{t}_L \gamma^{\mu} X_{\mu} s_L +
\cos\theta\overline{t}_L \gamma^{\mu} X_{\mu}\nu_{\tau L} +
\overline{t}_R \gamma^{\mu} X_{\mu} s_R +
\overline{b} \gamma^{\mu} X_{\mu} \tau) + {\rm H.c.}
\label{LX}
\end{equation}
The $Z'$ field is a linear combination of the gauge bosons
associated with $I_R$ and $B-L$. The orthogonal field
$B$ couples to weak hypercharge $Y$. The interaction
Lagrangian is
\begin{eqnarray}
{\cal L}_{Z',B} & = &
\frac{1}{\sqrt{g_R^2 + \frac{3}{2}g_s^2}} \sum_{\psi}
\overline{\psi} \Bigg( \gamma^{\mu} Z'_{\mu}
\left[ g_R^2 I_R P_R - \frac{3}{4} g_s^2 (B-L) \right]
\nonumber\\
& + & \gamma^{\mu} B_{\mu} \sqrt{\frac{3}{2}}g_R g_s
\left[ I_R P_R + \frac{B-L}{2} \right] \Bigg) \psi,
\label{LZ'B}
\end{eqnarray}
where $\psi = t,b,\tau,\nu$ and $P_R \equiv (1+\gamma_5)/2$.
The coupling constant for $B$ is identified with
$g_L\tan\theta_W$, where $g_L$ is the usual SU(2)$_L$
coupling constant. This allows us to calculate $g_R$ in
terms of the measured values of $g_L$, $\cos\theta_W$ and
$g_s$.

When $\Phi$ develops a nonzero VEV, $B$ and the neutral
gauge boson of SU(2)$_L$ form into the massless photon
and the massive $Z$ boson. The latter also mixes with $Z'$.
The left-sector $W$ boson acquires its standard mass
$m_{W_L}^2 = g_L (u_1^2 + u_2^2)/2$, and
it also mixes with the right-sector $W_R$.

I will also need to display the Yukawa couplings of
both the physical and unphysical Higgs bosons. Writing
\begin{equation}
\Phi = \left( \begin{array}{cc}
\phi_1^0\ & \phi_2^+\ \\ \phi_1^-\ & \phi_2^0\
\end{array} \right)
\end{equation}
the electroweak Yukawa Lagrangian is rewritten as
\begin{eqnarray}
{\cal L}_{\rm Yuk} & = &
\lambda_1 ( \overline{t}_L t_R \phi_1^0
+ \overline{t}_L b_R \phi_2^+ + \overline{b}_L t_R \phi_1^-
+ \overline{b}_L b_R \phi_2^0 \nonumber\\
& \ & \qquad + \ \overline{\nu}_L \nu_R \phi_1^0 +
\overline{\nu}_L \tau_R \phi_2^+
+ \overline{\tau}_L \nu_R \phi_1^- +
\overline{\tau}_L \tau_R \phi_2^0)\nonumber\\
& + & \lambda_2 ( \overline{t}_L t_R \phi_2^{0*}
- \overline{t}_L b_R \phi_1^+ - \overline{b}_L t_R \phi_2^-
+ \overline{b}_L b_R \phi_1^{0*} \nonumber\\
& \ & \qquad + \ \overline{\nu}_L \nu_R \phi_2^{0*} -
\overline{\nu}_L \tau_R \phi_1^+
- \overline{\tau}_L \nu_R \phi_2^- +
\overline{\tau}_L \tau_R \phi_1^{0*}) + {\rm H.c.}
\label{ewYuk}
\end{eqnarray}
Then writing
\begin{equation}
\chi = \left( \begin{array}{cc}
\chi^u\ & \chi^0\ \\ \chi^d\ & \chi^-\
\end{array} \right)
\end{equation}
I find that the non-electroweak Yukawa Lagrangian is
\begin{equation}
{\cal L}_{\rm Yuk} = n( \overline{N}_L t_R \chi^{u\dagger}
+ \overline{N}_L b_R \chi^{d\dagger}+
\overline{N}_L \tau_R \chi^+ + \overline{N}_L s_R \chi^{0*})
+ {\rm H.c.},
\end{equation}
where $\chi^u$ and $\chi^d$ are $1 \times 3$ row matrices
denoting the three colour components of these fields.

I now describe the gastronomy of the model.
The field $\chi^u$ is eaten by the $X$ boson, while $\chi^d$
is a physical colour triplet Higgs boson. In the limit that
$v \gg u_1, u_2$, the field $\chi^-$ is eaten by $W_R^-$,
while
\begin{equation}
g^- \equiv \cos\omega \phi_1^- - \sin\omega \phi_2^-
\end{equation}
where $\tan\omega \equiv u_2/u_1$
is eaten by $W_L^-$. The orthogonal field
\begin{equation}
H^- \equiv \sin\omega \phi_1^- + \cos\omega\phi_2^-
\end{equation}
is a physical charged
Higgs boson. For the case where spontaneous CP-violation
does not occur,
the real components of $\phi_1^0$, $\phi_2^0$
and $\chi^0$ mix to yield three physical fields. Two of the imaginary
components are eaten by the $Z'$ and $Z$. In the limit
$v \gg u_1, u_2$, the imaginary component of $\chi^0$ is
eaten by the $Z'$, while
$\sqrt{2}[\cos\omega{\rm Im}(\phi_1^0)
+ \sin\omega{\rm Im}(\phi_2^0)]$
is eaten by the $Z$, leaving the orthogonal field as
a physical CP-odd neutral Higgs boson. I will need the
interaction Lagrangian between $g^-$, $H^-$ and the
fermions. It is
\begin{eqnarray}
{\cal L}_{\rm Yuk}^+ & = &
a_g \overline{t}_L b_R g^+ + b_g \overline{b}_L t_R g^-
+ a_g \cos\theta \overline{\nu}_{\tau L} \tau_R g^+
+ a_g \sin\theta \overline{s}_L \tau_R g^+
+ b_g \overline{\tau}_L s_R g^-\nonumber\\
& + &
a_H \overline{t}_L b_R H^+ + b_H \overline{b}_L t_R H^-
+ a_H \cos\theta \overline{\nu}_{\tau L} \tau_R H^+
+ a_H \sin\theta \overline{s}_L \tau_R H^+\nonumber\\
& + & b_H \overline{\tau}_L s_R H^- + {\rm H.c.}
\end{eqnarray}
where
\begin{eqnarray}
a_g & \equiv & - \frac{m}{\sqrt{u_1^2 + u_2^2}},
\nonumber\\
b_g & \equiv & \frac{m_t}{\sqrt{u_1^2 + u_2^2}},
\nonumber\\
a_H & \equiv & \frac{1}{\cos 2\omega}
\frac{m_t - m\sin 2\omega}{\sqrt{u_1^2 + u_2^2}},
\nonumber\\
b_H & \equiv & \frac{1}{\cos 2\omega}
\frac{m_t \sin 2\omega - m}{\sqrt{u_1^2 + u_2^2}}
\end{eqnarray}
as can be easily seen from Eq.~(\ref{ewYuk}). The quantity $m$ is
the common tree-level mass for $b$ and $\tau$.

The primary task now is to discuss how radiative effects
modify the tree-level relation $m_b = m_{\tau}$. Before
doing so, I will make a brief comment about a cosmological
implication of the model. Because the unbroken symmetry
group contains no U(1) factors while the broken group
does, monopoles will be created during the $G_{PS}$
symmetry breaking phase transition in the early universe.
However, a simple calculation shows that monopoles
produced at a temperature of 1000 TeV are cosmologically
innocuous \cite{kt}. The number density of monopoles $n_M$ in the
visible universe today depends on how many causally
disconnected regions at $T = 1000$ TeV made up the spacetime that
subsequently evolved into the present day visible universe.
A rough order of magnitude estimate shows that
$n_M/s \sim (1000\ {\rm TeV}/M_{\rm Planck})^3$ where
$s$ is entropy density at the time of
monopole creation. If there is negligible monopole
annihilation then this ratio should remain roughly
constant. Using this to calculate the
fraction of critical density existing as monopoles I
find $\rho_M/\rho_{\rm cr} \sim
10^{14} (n_M/s) (m_M/10^3 {\rm TeV})$ where $m_M$ is
the monopole mass and is roughly 1000 TeV. Because
1000 TeV is much smaller than
$M_{\rm Planck} \sim 10^{16}$ TeV, I find that
$\rho_M/\rho_{\rm cr} \sim 10^{-26}$. I conclude that
looking for relic monopoles would be a very bad way to
test for a low-scale Pati-Salam symmetry breaking
phase transition.

\vskip 5 mm

\centerline{\large \bf 2.2
Renormalisation and $m_b/m_{\tau}$.}

\vskip 2 mm

The tree-level relation $m_b/m_{\tau}= 1$ holds at the
Pati-Salam symmetry breaking scale, which I will take to be about
$1000$ TeV. If radiative corrections due to threshold
effects at either the high symmetry breaking scale or the low
electroweak scale are ignored, then the change in this
ratio can be summarised by renormalisation group evolution. This means
that the renormalisation group equations are integrated
from $1000$ TeV to
the $b$ and $\tau$ mass scale of a few GeV \cite{ramond} using the
boundary condition $m_b = m_{\tau}$ at 1000 TeV. The
result of this evolution is that
\begin{equation}
m_b(m_b) = 4.11\ {\rm GeV}
\end{equation}
having chosen $m_{\tau}$ to come out correctly.
(A top mass of 174 GeV was used to derive this.) This
would be a very pleasing result if it could be
believed. It would mean that Pati-Salam theory
predicts the correct $b$ mass provided the symmetry
breaking scale is not too different from 1000 TeV.
Scales lower than 1000 TeV are phenomenologically
disallowed, and they seemingly predict too
small a value for $m_b$ anyway. Scales much higher
than 1000 TeV generate an
overweight bottom. Therefore the theory would predict
that observation of the rare decays $K_L^0 \to \mu^{\pm}e^{\mp}$
should occur in the not too distant future, as it is
precisely these decays that set the lower limit of about
1000 TeV on $m_X$ \cite{willenbrock}.
Furthermore, these decays seem to be
the most sensitive probe of the Pati-Salam model, so
no other rare decays should be observed during this
same time scale. The model could therefore either be
ruled out, or dramatic evidence gathered in its favour.

However, radiative corrections due to threshold effects {\it can be
extremely important} for a reason I now discuss. (This class of radiative
correction is not taken care of through renormalisation group
evolution.) The point is that some of the threshold corrections to
$m_{\tau} - m_b$ can be proportional to a large mass in the theory,
rather than $m_b$ or $m_{\tau}$ itself. In the present theory, the top
quark and the heavy neutrino mass eigenstates are all very massive
particles. It will turn out that charged Higgs boson graphs
produce a high mass scale threshold
correction in this theory
that is proportional to the top quark mass. Note
that a top quark mass of, say, 180 GeV will completely counteract the
$1/16\pi^2$ loop suppression factor.

I now identify those 1-loop self-energy graphs that
contribute to $m_b-m_{\tau}$. These are displayed
in Figs.1-7. Figure 1 shows the contributions from
the neutral gauge bosons in the model (the photon,
the gluons, the $Z$, and the $Z'$) together with that due
to the coloured gauge particle $X$. Figures 2 and 3 display
the contributions due to the electroweak charged
Higgs bosons $H^-$ and $g^-$ (I will work in an unphysical
gauge). Figures 4 and 5 contain the graphs involving
the charged $W$ bosons in both the left- and right-handed
sectors, while Fig.6 features graphs containing
components of $\chi$. Lastly, Fig.7 assembles all the
graphs that arise through mixing between the light
and heavy sectors of the theory.

It is sensible to group the graphs in the above
manner because of the way the divergences
cancel to give a finite $m_{\tau}-m_b$. All of the
individual graphs in Figs.1-6 are logarithmically
divergent\footnote{Actually they are superficially
linearly divergent, but the linear part is zero.},
but these divergences cancel within each class of
diagrams depicted in the separate figures. The
graphs in Fig.7 are all separately finite.

The quantity $m_b - m_{\tau}$ will now be calculated using
these graphs. The charged Higgs boson
graphs of Fig.2 will be of most interest.
However, I will first discuss
the evaluation of the set of
graphs in Fig.1 in detail, since this will illuminate how threshold
corrections and large logarithmic corrections associated with the
renormalisation group coexist. This calculation will also demonstrate
the relative unimportance of threshold corrections that are not
proportional to a large mass. Following this, I evaluate the important
threshold corrections arising from Fig.2. The Appendix provides full
details of these evaluations, together with a summary of the
contributions from Figs.3-7.

The result for Fig.1 is given by Eq.~(\ref{ans}) of the
Appendix which I reproduce here for convenience:
\begin{equation}
m_{\tau}-m_b|_{G} \simeq - m\frac{\alpha_s}{16\pi} \left(
3 \frac{2m_{Z'}^2 + 5 m_X^2}{m_{Z'}^2} \ln\frac{m_{Z'}^2}{m^2}
+ 12 \ln\frac{m_X^2}{m^2_{Z'}}
+ \frac{3}{2}\frac{2m_{Z'}^2 + 5m_X^2}{m_{Z'}^2}\right).
\label{result}
\end{equation}
This expression contains both a large logarithm
$\ln(m_{Z'}^2/m^2)$, which depends on the hierarchy between the
Pati-Salam and electroweak breaking scales, and additional pieces which
depend only on mass ratios involving the high mass sector. The large
logarithm is associated with those radiative corrections which can be
accounted for using the renormalisation group. The additional terms are
the sought after threshold corrections.

Let me discuss this distinction a little further: The set of graphs in
Fig.1 produce a finite correction to $m_{\tau}-m_b$; the logarithmic
divergences of the individual graphs exactly cancel between the graphs.
Since the cancellation occurs
between graphs containing light gauge bosons and
those containing heavy gauge bosons, there emerges by necessity a
large logarithm. If only the light gauge bosons of the SM were included,
then $m_{\tau}-m_b$ would diverge. However, because the
heavy sector of the theory ``knows'' about the physics which is trying
to maintain $m_{\tau}=m_b$, the heavy gauge boson
graphs effectively act as an ulraviolet regulator for the logarithmic
divergence produced by the light gauge boson graphs. The logarithmic
divergence is turned into a large logarithm. The presence of this large
dimensionless quantity calls into question the usefulness of 1-loop
perturbation theory, because the effective expansion parameter is not
the square of a coupling constant but rather the square of a coupling
constant multiplied by the large logarithm. This means that higher order
graphs may well provide numerically important corrections to the 1-loop
expression. The task of calculating these corrections can, fortunately,
be elegantly performed by solving the renormalisation group equations,
a process that is tantamount to summing these large
logarithms to all orders.

I therefore simply omit the large logarithmic term obtained from Fig.1,
knowing that its effects will be incorporated by solving the
renormalisation group equations. The remaining terms, however, cannot be
accounted for in this manner. These threshold corrections, so-called
because they depend on heavy mass ratios only, can be viewed as setting
up the boundary condition on $m_{\tau}-m_b$ at the Pati-Salam
breaking scale that one must use to solve
the renormalisation group equations.

Note that there is an ambiguity in
how to separate the large logarithmic term from the threshold
corrections. Should the large mass in the logarithm be $m_{Z'}$ as shown
above, or $m_X$ instead? In other words, should the running start from
the mass $m_{Z'}$ or the mass $m_X$?
This ambiguity will not be numerically important in this paper,
because the large threshold corrections I will obtain
from Fig.2 will not need to be separated from a large logarithmic term.

Let us now obtain a numerical estimate for the size of the threshold
corrections. They depend through the heavy mass ratios on the coupling
constants of SU(3)$_c$ and SU(2)$_R$ (the VEV of $\chi$ cancels out).
Renormalisation evolution for $\alpha_s$ shows that
\begin{equation}
\alpha_s(\Lambda) =
\frac{\alpha_s(m_Z)}{1 + \frac{7}{2\pi} \alpha_s(m_Z)
\ln(\Lambda/m_Z)}.
\end{equation}
Inputting $\alpha_s(m_Z) = 0.118$ produces
\begin{equation}
\alpha_s(1000\ {\rm TeV}) = 0.053.
\end{equation}
The right-handed SU(2) coupling constant is given by
\begin{equation}
\frac{1}{\alpha_R} = \frac{1}{\alpha_Y} -
\frac{2}{3\alpha_3},
\end{equation}
and renormalisation group evolution implies that
\begin{equation}
\alpha_R(\Lambda) =
\frac{3\alpha_Y(m_Z)\alpha_s(m_Z)}
{3\alpha_s(m_Z)- 2\alpha_Y(m_Z) -
\frac{35}{2\pi}\alpha_Y(m_Z)\alpha_s(m_Z)\ln(\Lambda/m_Z)}.
\end{equation}
Using $\alpha_Y(m_Z) = 0.0101$ yields
\begin{equation}
\alpha_R(1000\ {\rm TeV}) = 0.013.
\end{equation}
Inputting these values into the last two terms of Eqn.~(\ref{result})
shows that the threshold corrections
produce $m_{\tau}-m_b \simeq 10$'s of
MeV. Since renormalisation group evolution alters this
quantity by a few GeV, these threshold terms can be safely neglected.

However, the graphs of Fig.2 produce much larger threshold corrections
due to the presence of the top quark in the loop and the top-quark mass
in the vertices involving the physical charged Higgs boson. Note first
of all that it is natural to take the mass $m_H$ of $H^-$ to be of the
order of the Pati-Salam breaking scale. The point is that the
linear combination that contains $H^-$ of the two SU(2)$_L$
doublets embedded in $\Phi$ has zero VEV. This linear combination
therefore plays no role in setting the scale of electroweak
symmetry breakdown, and the masses of the component fields may
take on ``natural'' values of the order of the high symmetry breaking
scale. This is phenomenologically useful because it means that the
effective neutral flavour-changing effects that $H^-$ produces at 1-loop
order and above are very suppressed \cite{babu}.
Furthermore, it is clear that no
large logarithm will arise for these graphs because they do not separate
into a SM subset and a Pati-Salam subset that cancel each others
logarithmic divergences.

The physical charged Higgs boson graphs in Fig.2 yield
\begin{equation}
m_{\tau}-m_b|_{H} \simeq - \frac{1}{16\pi^2}
\frac{m_s^2 - m_t^2}{m_H^2 - m_s^2}
\frac{m_t(m_t - m\sin2\omega)(m_t\sin2\omega - m)}
{(u_1^2 + u_2^2)\cos^22\omega}
\ln\left(\frac{m_s^2}{m^2_H}\right)
\label{result_H}
\end{equation}
in the limit that $m_s, m_H \gg m_t$. I have also assumed in the
approximate expression given above that there is no accidental
cancellation between $m_t\sin2\omega$ and $m$. This
threshold correction can clearly produce a mass difference between
$m_{\tau}$ and $m_b$ of the order of a GeV, provided this accidental
cancellation does not occur. The ``common'' mass $m$ of $\tau$ and
$b$ at the Pati-Salam breaking scale must be about the same as the
measured $m_{\tau}$, namely about $1.8$ GeV, because $m_{\tau}$ does not
evolve strongly under the renormalisation group. The above threshold
effect can therefore alter the initial ratio $m_b/m_{\tau}$ by up to
$50\%$. This correction is thus as numerically
significant as those incorporated through the renormalisation group.
The sign of the correction depends on the unknown parameter $\omega$,
and therefore cannot be predicted. It can either raise
or lower the mass ratio by up to $50\%$. Interestingly, the sign does
not depend on which of $m_s$ and $m_H$ is larger (although the magnitude
of the correction is strongly dependent on these masses).

\vskip 5 mm

\centerline{\large \bf 2.3 Discussion}

\vskip 2 mm

The calculation demonstrates that generally speaking
one must take care in the use of
renormalisation group evolution to predict low-energy masses. It is
quite possible for low-energy masses to be very sensitive to unknown
details surrounding the high symmetry breaking sector, through threshold
corrections that are enhanced by a large mass. In the particular model
I analysed, the large threshold corrections were produced by graphs
involving the physical charged Higgs boson only. It is possible that
most models lacking such a particle will also lack large threshold
corrections. For instance, one may choose to gauge only the U(1) subgroup
of SU(2)$_R$ rather than whole right-handed weak-isospin group. One
could then try to construct a model with a single electroweak Higgs
doublet rather than the two doublets that are effectively contained
within $\Phi$. A physical charged Higgs boson would then be absent, and
perhaps also large threshold effects.

It is interesting that the sign of the large threshold correction
depends crucially on $\omega$ which in turn depends on the relative sign
between the two electroweak VEVs $u_1$ and $u_2$. If the correction
produces $m_b > m_{\tau}$ at 1000 TeV, then renormalisation group
evolution will produce on overly massive bottom quark. This would
necessitate that the accidental cancellation between $m_t\sin2\omega$
and $m$ occur to some extent. If the correction produces $m_b <
m_{\tau}$, then the masses will need to be evolved for a
longer period in order to produce a phenomenologically acceptable outcome.
This would mean that the Pati-Salam breaking scale should be higher than
the nominal value of 1000 TeV that I have been considering.

It would be interesting to extend this analysis to a three
generation model. Are radiative corrections in the three generation
of the model able to accomodate $s$-$\mu$
and $d$-$e$ mass splitting? This may be possible, given enough freedom
to combine renormalisation group evolution and potentially large threshold
corrections. It is, however, not obvious that this will work because
one would generically expect Higgs boson effects to be less
important for lower generations.

However, it is perhaps more worthwhile to think of
some horizontal structure that may increase the
predictivity of the model. A question in this context is whether or
not it would be interesting to invoke a Georgi-Jarlskog
texture via a $(15,2,2)$ Higgs boson \cite{gj},
or whether such a tree-level
texture would be wiped out by radiative corrections. The important issue
of predictivity also raises the question of how to reduce the freedom
one has in moulding the size of threshold corrections by unknown details
of the heavy sector of the theory. It would clearly be interesting to
construct the heavy sector in the simplest possible manner in order to
reduce the number of experimentally unknowable parameters.

\vskip 1 cm

\centerline{\Large \bf 3. Towards $t$-$b$-$\tau$ unification}

\vskip 3 mm

As mentioned in the previous section, if the electroweak
bidoublet $\Phi$ is chosen to be real then mass
equality between isospin partners occurs at
tree-level. With $\Phi = \Phi^c$ we have that
\begin{equation}
\Phi = \left( \begin{array}{cc}
\phi^0\ & -\phi^+\ \\ \phi^-\ & \ \phi^{0*}
\end{array} \right)
\end{equation}
and the Yukawa Lagrangian
\begin{equation}
{\cal L}_{\rm Yuk} = \lambda
{\rm Tr}(\overline{f}_L \Phi f_R) + {\rm H.c.}
\end{equation}
produces
\begin{equation}
m_t = m_b = m_{\tau} = m_{\nu}^{\rm Dirac} = \lambda u,
\end{equation}
having used
\begin{equation}
\langle\Phi\rangle =
\left( \begin{array}{cc}
u\ & 0\ \\ 0\ & u\
\end{array} \right).
\end{equation}
The full power of $G_{PS}$ to relate masses is thus
evident. A useful way to view the above phenomenon is that
custodial SU(2) has been gauged and upgraded to an exact
symmetry of the Lagrangian by its identification with SU(2)$_R$.

I have demonstrated that radiative corrections
can alter mass ratios dramatically. However, the measured
ratio $m_t/m_{\tau}$ is about $100$ and thus threshold
corrections cannot plausibly be used to
fix up $m_t = m_{\tau}$, unless the large mass used to enhance the
correction is very much larger than $m_t$ \cite{pati}. One may speculate
that the neutrino sector of a theory may produce such an effect,
although this did not happen in the Pati-Salam model considered in the
previous Section.

The obvious alternative is to use some form of see-saw
mechanism to depress $m_{\tau}$ and $m_b$ relative to
$m_t$, just as one may do in the neutrino sector. In other
words, mixing effects rather than
radiative corrections can be relied upon
to explain why $m_{\nu_{\tau}}
\ll m_{\tau,b} \ll m_t$, while radiative
corrections only are used to accomodate the ratio $m_b/m_{\tau}$.

It is therefore rather interesting to observe that the
${\bf 10}$ of SU(4) has the branching rule
\begin{equation}
{\bf 10} \to {\bf 6}(\frac{2}{3}) \oplus
{\bf 3}(-\frac{2}{3}) \oplus {\bf 1}(-2)
\end{equation}
to SU(3)$\otimes$U(1)$_{B-L}$. The colour triplet
component has electric charge $-1/3$, while the colour
singlet has electric charge $-1$. Within this one
irredicible representation lie the correct states
that can mix with $b$ and $\tau$ in a
see-saw manner. Furthermore, the electric charge
$+2/3$ state is absent. One can therefore arrange for
$m_b$ and $m_{\tau}$ to be lowered with respect to $m_t$.
In addition, a fermion in the $(10,1,1)$
representation of $G_{PS}$ can mix with $f_R$
via Yukawa coupling with $\chi$. All the ingrediants
are there within the group theory of SU(4) to
do exactly what I want to do. I find this to be
a rather striking fact.

So, I write down a new Pati-Salam model that contains
the fermions
\begin{equation}
f_L \sim (4,2,1),\ \ f_R \sim (4,1,2),\ \
F_L \sim (10,1,1),\ \ F_R \sim (10,1,1),\ \
N_L \sim (1,1,1)
\end{equation}
and the Higgs bosons
\begin{equation}
\Phi = \Phi^c \sim (1,2,2)\quad {\rm and}\quad
\chi \sim (4,1,2).
\end{equation}
The full Yukawa Lagrangian is
\begin{equation}
{\cal L}_{\rm Yuk} =
\lambda {\rm Tr}(\overline{f}_L \Phi f_R)
+ h {\rm Tr}(\overline{F}_L \chi^T i\tau_2 f_R)
+ n \overline{N}_L {\rm Tr}(\chi^{\dagger}f_R)
+ M_F {\rm Tr}(\overline{F}_L F_R) + {\rm H.c.}
\end{equation}
where $F_{L,R}$ have been written as symmetric
$4\times 4$ matrices which undergo the SU(4)
transformation $F_{L,R} \to U_4 F_{L,R} U_4^T$.
In component form,
\begin{equation}
F = \left( \begin{array}{cc}
S\ & \frac{B}{\sqrt{2}}\ \\ \frac{B^T}{\sqrt{2}}\ & E\
\end{array} \right)
\label{Fcomps}
\end{equation}
where $S$ is a $3\times 3$ symmetric matrix
representing the colour sextet, $B$ is a $3\times 1$
column matrix representing the colour triplet and
$E$ is the colour singlet. The $\sqrt{2}$ in this equation
is required in order to normalise the kinetic energy
terms for $B$ and $E$ consistently.

The top and Dirac neutrino masses are simply
\begin{equation}
m_t = m_{\nu}^{\rm Dirac} = \lambda u.
\end{equation}
However, bottom and tau now have $2\times 2$
mass matrices given by
\begin{equation}
{\cal L}_{\rm b} =
\left( \begin{array}{cc}
\overline{b}_L\ & \overline{B}_L
\end{array} \right)
\left( \begin{array}{cc}
m_t\ & 0\ \\ m_B\ & M_F\
\end{array} \right)
\left( \begin{array}{c}
b_R \\ B_R \end{array} \right) + {\rm H.c.}
\end{equation}
and
\begin{equation}
{\cal L}_{\rm \tau} =
\left( \begin{array}{cc}
\overline{\tau}_L\ & \overline{E}_L
\end{array} \right)
\left( \begin{array}{cc}
m_t\ \ & 0\ \\ \sqrt{2}m_B\ & M_F\
\end{array} \right)
\left( \begin{array}{c}
\tau_R \\ E_R \end{array} \right) + {\rm H.c.}
\end{equation}
where $m_B \equiv hv/\sqrt{2}$.
The $\sqrt{2}$ in the $\tau$ mass matrix comes from
the $\sqrt{2}$ in Eq.~(\ref{Fcomps}).

Since $v \gg u$, we
expect that $m_B \gg m_t$, unless the Yukawa coupling
constant $h$ is very small.
One large eigenvalue and one small eigenvalue is thus expected
from each mass matrix, provided the bare
mass $M_F$ is not too large. In fact, if $M_F \ll m_B$
(but not necessarily small compared to $m_t$)
the smallest eigenvalues are roughly
$\sqrt{2}m_t M_F/m_B$ for
the $b$ system, and $m_t M_F/m_B$ for the $\tau$-system.
This shows that mixing between $f$ and $F$ can indeed
suppress $m_b$ and $m_{\tau}$ with respect to $m_t$.
So, the small eigenvalues are identified
with $m_b$ and $m_{\tau}$, while I
will call the large eigenvalues $m_{b'}$ and
$m_{\tau'}$.

The two mass matrices produce four eigenvalues
in terms of three parameters. This means there is
one relation connecting them.
The relation can be written most usefully in the form
\begin{equation}
\frac{m_b}{m_{\tau}} =
\left[
\frac{2 - \frac{m_t^2}{m_{b'}^2}
- \frac{m_{\tau}^2}{m_{b'}^2}}
{1 + \frac{m_{\tau}^2}{m_t^2}
- 2\frac{m_{\tau}^2}{m_{b'}^2}} \right]^{\frac{1}{2}}
\end{equation}
where I have chosen $m_t$ rather than $m_{\tau'}$
as one of the mass parameters on the right-hand side.
(Note that $m_{\tau'} = m_{b'}m_b/m_{\tau}$.) Since
$m_{\tau} \ll m_t,\ m_{b'}$ is required,
\begin{equation}
\frac{m_b}{m_{\tau}} \simeq
\sqrt{2 - \frac{m_t^2}{m_{b'}^2}}
\end{equation}
must hold so that $m_b \to \sqrt{2} m_{\tau}$ as
$m_{b'} \to \infty$.

For the interesting case where $m_t \ll M_F \ll m_B$,
the light mass eigenstate fields $\tilde{b}$
and $\tilde{\tau}$ are
\begin{equation}
\tilde{b}_L \simeq b_L - \frac{m_t}{m_B} B_L,
\quad \tilde{b}_R \simeq B_R - \frac{M_F}{m_B} b_R
\end{equation}
and
\begin{equation}
\tilde{\tau}_L \simeq \tau_L
- \frac{m_t}{\sqrt{2}m_B} E_L,
\quad \tilde{\tau}_R \simeq E_R
- \frac{M_F}{\sqrt{2}m_B} \tau_R.
\end{equation}
Thus the left-handed mass eigenstates $b$ and
$\tau$ are predominantly in $f_L$, while their
right-handed projections are mostly in $F_R$. This is
important because it means the light mass eigenstates
will feel the standard left-handed weak interactions
to a higher degree of accuracy, as is phenomenologically
required. The right-handed states will, however, have
their couplings to right-sector weak bosons
suppressed by $M_F/m_B$. This behaviour is similar to
Ma's alternative formulation of left-right symmetry \cite{ma}.
Because $m_b \ll M_F, m_B$ is phenomenologiclly necessary,
$m_b \simeq \sqrt{2} m_{\tau}$
must hold to a good level of approximation at tree-level.

So, I have shown that mixing effects can induce the
pattern $m_{\nu_{\tau}} = 0 \ll m_b, m_{\tau} \ll m_t$
provided $M_F$ is not too large. (The neutrino sector here
is identical to that of the Sec.II.) It remains to be
seen whether or not radiative effects can provide a
successful value for $m_b/m_{\tau}$. The explicit calculation
of the necessary diagrams is beyond the scope of this paper,
although experience with the previous model suggests that
there may be large threshold corrections due to Higgs
boson graphs that
can be arranged to produce a phenomenological successful
mass pattern for the third family, particularly given the involvement
of the heavy fermions in some relevant diagrams.
It may be that the additional
factor of roughly $\sqrt{2}$ in the tree-level
value of $m_b/m_{\tau}$ can be negated by a threshold
correction, with the ensuing boundary condition $m_{\tau} \simeq m_b$ at
1000 TeV then producing successful low-energy values.

\vskip 5 mm

\centerline{\Large \bf 4. Conclusion}

\vskip 3 mm

The idea that Pati-Salam SU(4) might be broken at a relatively
low energy such as 1000 TeV is a very appealing one. I have
shown in this work how the model ought to be constructed in
order to achieve this in a way consistent with Hot Big Bang
cosmology and particle phenomenology. I pointed out that a
different and simpler Higgs sector to that usually employed
to break SU(4) is required. The simplest version of this model
predicts massless neutrinos, although massive neutrinos are not
difficult to incorporate.

The core of the paper was then a calculation of the
radiative corrections to the tree-level mass relation
$m_b = m_{\tau}$ induced by mass thresholds.
I found that the set of graphs involving the charged Higgs boson
produces a generically large correction, enhanced by $m_t/m_{\tau}$.
This can alter the ratio $m_b/m_{\tau}$ by up to about $50\%$.
Whether this correction increases or decreases
the ratio depends on the relative sign between the two VEVs that
break the electroweak group. If the ratio is increased, then the
combined effect of the threshold correction and renormalisation group
evolution tends to produce an overly massive bottom quark. If the ratio
is decreased, then the scale of Pati-Salam symmetry breaking needs to be
raised in order to allow the masses to run for longer under the
renormalisation group. In either case, the generically large
threshold correction can be reduced by a fine-tuning of parameters.

It was then demonstrated that the identification of SU(2)$_R$ with
custodial SU(2) can yield $t$-$b$-$\tau$ unification at
tree-level when combined with Pati-Salam SU(4). I showed how the
hierarchy $m_{\nu_{\tau}} \ll m_{b,\tau} \ll m_t$ can
arise due to two different see-saw mechanisms, and I conjectured
that the $b$-$\tau$ splitting can possibly be accomodated
within the theory.

I am therefore able to reach the important conclusion that
the observed mass pattern of the third generation of quarks and
leptons can be reproduced by a Pati-Salam SU(4) theory
far below a hypothetical GUT scale. This scale could be just above the
current lower bound of about 1000 TeV. However, one must be mindful that
large threshold corrections be incorporated (or cancelled off as the case
might be), as well as renormalisation
group effects.
This motivates that an on-going search for rare
processes such as $K^0_L \to \mu^{\pm} e^{\mp}$ be maintained. The
detection of such a process may provide the first experimental
clue to the physics behind the fermion mass problem
and the relationship between quarks and leptons.

\vskip 1 cm

\centerline{\bf Note Added}

\vskip 5 mm

After these calculations were substantially complete, a somewhat
similar model was considered in Ref.\cite{worah}. It was shown here
that threshold corrections can induce mass corrections of the
order of several GeV, which lends further support to the idea that a
combination of renormalisation group evolution and large threshold
corrections may be interesting for the fermion mass problem
in theories with new physics far below $10^{16}$ GeV. Although this
paper explicitly considers a GUT-scale theory, the effects found
can also occur in lower scale physics, as was noted in the manuscript.

\vskip 2 cm

\centerline{\Large \bf Acknowledgements}

\vskip 5 mm

The author is indebted to K. S. Babu for his collaboration on
portions of this work, and in particular for checking some of the
calculations. (Any errors in the manuscript are the
sole responsibility of the author, however).
He would like to thank Professor J. C. Taylor and the
theoretical particle physics group in the Department of Applied
Mathematics and Theoretical Physics at the University of Cambridge
for kind hospitality while this work was performed. He would also
like to thank R. B. Mann and R. Thorne for some brief but useful
discussions, and R. Foot for a helpful piece of
correspondence. This work was partially supported by the Australian
Research Council and partially by the University of Melbourne.

\newpage

\centerline{\Large \bf Appendix}

\vskip 3 mm

In this Appendix I will calculate the graphs
displayed in Figs.1-7, working in Feynman gauge for
all of the gauge interactions. A highly non-trivial
consistency check on the
calculation will be that all of the divergences
should cancel in $m_{\tau}-m_b$.

A pragmatic approach to the regularisation
of the various integrals will be adopted, employing either
dimensional regularisation or Pauli-Villars
regularisation depending on what happens to
be convenient. Since I am calculating a finite quantity,
no inconsistency is introduced by employing two
different regularisation procedures.

\vskip 3 mm

\centerline{\large \bf A.1 Graphs in Figure 1}

\vskip 2 mm

In this first subsection I will calculate the
contribution of the diagrams in Fig.1
To simplify the task,
the mass of the $Z$-boson will be set to zero, thus making
it degenerate with the photon. Everything can then be rewritten
in terms of $B$ and
$W^0_L$, the latter being the neutral gauge boson
of SU(2)$_L$. But then the $W^0_L$ boson
graph need not be considered,
since it couples universally to $b$ and $\tau$. Since I am
interested in threshold corrections due to heavy sector masses, my
neglect of $m_Z$ will be of no numerical significance.

It is useful to first consider a general gauge
interaction of the form
\begin{equation}
{\cal L}(x,y) = \overline{f}_1 \gamma^{\mu}
(x + y\gamma_5) f_2 A_{\mu}
\end{equation}
where $f_{1,2}$ both have mass $m$, $A$ has mass M
and where $f_1$ and $f_2$ may be the same field.
The 1-loop self-energy generated by
${\cal L}(x,y)$ is
\begin{equation}
-i\Sigma_F = - \int \frac{d^4k}{(2\pi)^4}
\frac{\gamma^{\mu}(x +y\gamma_5)(\hat{p} + \hat{k} + m)
\gamma_{\mu}(x + y\gamma_5)}
{[(k + p)^2 - m^2](k^2 - M^2)},
\end{equation}
where the symbol $\hat{k}$ means $\gamma^{\mu}k_{\mu}$.

These terms contain both wave-function renormalisation
constants as well as mass shifts, and I seek only
the latter. A general fermion
self-energy $\Sigma$ may be written in the form
\begin{equation}
\Sigma = A(\hat{p}-m) + B (\hat{p}-m)\gamma_5
+ C \gamma_5 (\hat{p}-m) + \delta m,
\label{Sigma}
\end{equation}
where $A$, $B$ and $C$ contribute to wave-function
renormalisation while $\delta m$ is the mass shift.
The $\gamma_5$ dependence shown above is required
because of the complication that the gauge
interactions I consider are chiral. It is important
to realise that the coefficient of $\gamma_5$ in
the self-energy contributes only to wave-function
renormalisation. One might fear that this cannot be
the case because in general $\Sigma$ should
have a term of the form $\delta\mu\gamma_5$,
which looks like
a peculiar $\gamma_5$-dependent contribution to
the mass. However, the identity
\begin{equation}
\gamma_5 =
- \frac{(\hat{p}-m)\gamma_5 + \gamma_5(\hat{p}-m)}{2m}
\end{equation}
shows that such a term can always be subsumed into
the $B$ and $C$ terms in Eq.~(\ref{Sigma}). Since
these terms cannot shift the pole away from
$\hat{p}=m$, they do not contribute to mass
renormalisation. In practice then, the mass shift is
isolated by setting $\hat{p}=m$, $p^2 = m^2$ and dropping
the contribution proportional to $\gamma_5$.

To proceed I first
regularise the divergent integrals by continuing to
$n$-dimensions. Although $m_{\tau}-m_b$ will be a finite
quantity, it is the sum of integrals that are separately
divergent. In order to be certain that no
errors are introduced by a naive cancellation
of infinite quantities, I feel
it prudent to regularise the integrals first.\footnote{ By ``naive''
I mean the combining of the integrands of Feynman integrals
using a common denominator after having simplified the numerators
using 4-dimensional Dirac algebra.} This may seem like
pedantry because the answer turns out to be identical
to that obtained by just such a naive cancellation. However, I view
the cancellation of regularised divergences as a justification for
veracity of the naive method.

To avoid $n$-dimensional $\gamma$-matrix
algebra involving $\gamma_5$, the positions
of all the $\gamma$-matrices in the numerator are frozen. Since the
integral is now finite, all ordinary
manipulations except for Dirac algebra can be performed.
Equations \ref{LX} and \ref{LZ'B} are now used in
conjunction with the familiar gluon interaction with
quarks to obtain the $x$ and $y$ parameters for each
diagram. The contributions are then summed with the
appropriate colour factors for the $X$ boson and gluon
graphs inserted.

The self-energies for $\tau$ and $b$ are
\begin{equation}
-i\Sigma(f) = - \int \frac{d^nk}{(2\pi)^n}
\frac{N(f)}{(k + p)^2 - m^2}
\end{equation}
where $f=\tau,b$ and
\begin{eqnarray}
N(\tau) & = & \frac{3}{8}
\frac{g_R^2 g_s^2}{g_R^2 + \frac{3}{2}g_s^2}
\frac{\gamma^{\mu}(1+P_R)(\hat{p}+\hat{k}+m)
\gamma_{\mu}(1+P_R)}{k^2}\nonumber\\
& + & \frac{1}{16}\frac{1}{g_R^2 + \frac{3}{2}g_s^2}
\frac{\gamma^{\mu}(3g_s^2-2g_R^2 P_R)(\hat{p}+\hat{k}+m)
\gamma_{\mu}(3g_s^2-2g_R^2 P_R)}{k^2 - m_{Z'}^2}\nonumber\\
& + & \frac{3g_s^2}{2}
\frac{\gamma^{\mu}(\hat{p}+\hat{k}+m)\gamma_{\mu}}
{k^2 - m_X^2}.
\end{eqnarray}
The three terms in this equation come from the $B$ graph,
the $Z'$ graph and the $X$ graph,
respectively. The corresponding expression for $b$ is
\begin{eqnarray}
N(b) & = & \frac{1}{24}
\frac{g_R^2 g_s^2}{g_R^2 + \frac{3}{2}g_s^2}
\frac{\gamma^{\mu}(1-3P_R)(\hat{p}+\hat{k}+m)
\gamma_{\mu}(1-3P_R)}{k^2}\nonumber\\
& + & \frac{1}{16}\frac{1}{g_R^2 + \frac{3}{2}g_s^2}
\frac{\gamma^{\mu}(g_s^2+2g_R^2 P_R)(\hat{p}+\hat{k}+m)
\gamma_{\mu}(g_s^2+2g_R^2 P_R)}{k^2 - m_{Z'}^2}\nonumber\\
& + & \frac{g_s^2}{2}
\frac{\gamma^{\mu}(\hat{p}+\hat{k}+m)\gamma_{\mu}}
{k^2 - m_X^2}\nonumber\\
& + &\frac{4g_s^2}{3}
\frac{\gamma^{\mu}(\hat{p}+\hat{k}+m)\gamma_{\mu}}{k^2},
\end{eqnarray}
where the fourth term is due to the gluon graph.
Expanding the numerators above, without commuting any
of the Dirac matrices through each other, and subtracting
the $b$ term from the $\tau$ term I find that
\begin{equation}
-i(\Sigma_{\tau}-\Sigma_{b}) =
- \frac{1}{g_R^2 + \frac{3}{2}g_s^2} \int
\frac{d^nk}{(2\pi)^n}
\frac{N}{(k + p)^2 - m^2}
\label{Sigmataub}
\end{equation}
where
\begin{eqnarray}
N & = &
\frac{g_R^2 g_s^2
[ \frac{1}{3} \gamma^{\mu}(\hat{p}+\hat{k}+m)\gamma_{\mu}
+ \frac{1}{2} \gamma^{\mu}P_R(\hat{p}+\hat{k}+m)\gamma_{\mu}
+ \frac{1}{2} \gamma^{\mu}(\hat{p}+\hat{k}+m)\gamma_{\mu}P_R]}
{k^2}\nonumber\\
& + &
\frac{ g_s^2 [
\frac{1}{2}g_s^2\gamma^{\mu}(\hat{p}+\hat{k}+m)\gamma_{\mu}
- \frac{1}{2}g_R^2
\gamma^{\mu}P_R (\hat{p}+\hat{k}+m)\gamma_{\mu}
- \frac{1}{2} g_R^2
\gamma^{\mu}(\hat{p}+\hat{k}+m)\gamma_{\mu}P_R]}
{k^2 - m_{Z'}^2}
\nonumber\\
& + & \frac{g_s^2(g_R^2 + \frac{3}{2}g_s^2)\gamma^{\mu}
(\hat{p}+\hat{k}+m)\gamma_{\mu}}{k^2 - m_X^2}\nonumber\\
& - & \frac{\frac{4}{3} g_s^2 (g_R^2 + \frac{3}{2}g_s^2)
\gamma^{\mu}(\hat{p}+\hat{k}+m)\gamma_{\mu}}{k^2}.
\end{eqnarray}
The cancellation of the divergences is evident in this
expression. The individually divergent pieces may be
isolated by temporarily setting $m_{Z'}=m_X=0$. The
terms containing $P_R$ cancel between the $B$ and
$Z'$ graphs, while all four graphs are required to
see the cancellation in the $P_R$-independent terms.
Since $-i(\Sigma_{\tau}-\Sigma_{b})$ is finite, the integral can
now be continued back to 4-dimensions and
Dirac algebra used.

This result illustrates the general phenomenon that
the heavy particles act effectively as ultraviolet
cutoffs for the self-energy graphs involving SM
particles only. If only the $B$ boson and gluon
graphs are included, then $-i(\Sigma_{\tau}-\Sigma_{b})$
is divergent. This is as
expected because the low-energy effective theory
is the SM which requires a counterterm to absorb
such a divergence. When all four graphs are included,
the full SU(4) symmetry of the underlying
Lagrangian is felt by $-i(\Sigma_{\tau}-\Sigma_{b})$
and it is revealed as a finite quantity.

Equation \ref{Sigmataub} may be rewritten more
compactly as
\begin{eqnarray}
-i(\Sigma_{\tau}-\Sigma_{b}) & = &
\frac{g_s^2}{2} (9m_X^2 - 2m_{Z'}^2)
\int \frac{d^4k}{(2\pi)^4}
\frac{\hat{p}+\hat{k}-2m}{D}
\nonumber\\
& - & \frac{g_s^2}{2} m_X^2 (5m_X^2 + 2m_{Z'}^2)
\int \frac{d^4k}{(2\pi)^4}
\frac{\hat{p}+\hat{k}-2m}{Dk^2} + (\gamma_5\ {\rm term}),
\label{tau-b}
\end{eqnarray}
where
\begin{equation}
D \equiv [(k + p)^2 - m^2](k^2 - m_{Z'}^2)(k^2 - m_X^2).
\end{equation}
The $\gamma_5$ term is now dropped, and
the remaining integrals have to be evaluated further to isolate
the mass shift.

The required integrals are
\begin{equation}
I_3 = \int\frac{d^4k}{(2\pi)^4} \frac{1}{D},\quad
I_4= \int\frac{d^4k}{(2\pi)^4} \frac{1}{Dk^2},
\end{equation}
and
\begin{equation}
\hat{I}_3 =  \int\frac{d^4k}{(2\pi)^4}
\frac{\hat{k}}{D},\quad
\hat{I}_4= \int\frac{d^4k}{(2\pi)^4} \frac{\hat{k}}{Dk^2}.
\end{equation}
I now approximately evaluate these integrals with $p^2 = m^2$
under the condition that $m_X^2 \sim m^2_{Z'} \gg m^2$.

The results are,
\begin{eqnarray}
I_3 & \simeq & \frac{i}{16\pi^2}\frac{1}{m^2_{Z'}-m_X^2}
\ln\left(\frac{m^2_X}{m^2_{Z'}}\right);\\
I_4 & \simeq & \frac{i}{16\pi^2}\frac{1}{m^2_X}\left[
\frac{1}{m^2_{Z'}}\ln\left(\frac{m^2_{Z'}}{m^2}\right)
+ \frac{1}{m^2_{Z'}} +
\frac{1}{m^2_{Z'}-m^2_X}\ln\left(\frac{m^2_X}{m^2_{Z'}}\right)
\right];\\
\hat{I}_3 & \simeq & -\frac{\hat{p}}{2}I_3;\\
\hat{I}_4 & \simeq & \frac{i}{32\pi^2}\frac{\hat{p}}{m^2_X}
\left[-\frac{1}{m^2_{Z'}}\ln\left(\frac{m^2_{Z'}}{m^2}\right)
+ \frac{1}{2m^2_{Z'}} + \frac{1}{m^2_X-m^2_{Z'}}
\ln\left(\frac{m^2_X}{m^2_{Z'}}\right)\right].
\end{eqnarray}
Note that $I_4$ and $\hat{I}_4$ contain the large logarithms associated
with the renormalisation group.

Substituting these expressions into Eqn.~(\ref{tau-b}) and replacing
$\hat{p}$ by $m$ to extract the mass shift part only, I find that
\begin{equation}
m_{\tau}-m_b|_{G} \simeq - m\frac{\alpha_s}{16\pi} \left(
3 \frac{2m_{Z'}^2 + 5 m_X^2}{m_{Z'}^2} \ln\frac{m_{Z'}^2}{m^2}
+ 12 \ln\frac{m_X^2}{m^2_{Z'}}
+ \frac{3}{2}\frac{2m_{Z'}^2 + 5m_X^2}{m_{Z'}^2}\right).
\label{ans}
\end{equation}
where I have kept only the large logarithmic terms followed by the
largest threshold corrections.

\vskip 3 mm

\centerline{\large \bf A.2 Graphs in Figure 2}

\vskip 2 mm

By contrast with the previous subsection, I will
not employ dimensional regularisation but rather
Pauli-Villars regularisation in this subsection.
This is convenient because all of the graphs
in Fig.2 have the same boson $H^-$ in the loop,
and so the Pauli-Villars cut-off $\Lambda$ is
necessarily the same for all the graphs. In
Fig.1 all of the bosons are different and therefore
in principle one could employ different cut-off
masses for each of the bosons. This would cloud
the issue of divergence cancellation between the
graphs, although it could still be demonstrated
in the limit that all of the regulating masses
were simultaneously large. Furthermore,
once the Pauli-Villars regulator is introduced
for the graphs in Fig.2 I am free to use
4-dimensional Dirac algebra immediately. This is
very convenient.\footnote{In fact, the calculations
show that you cannot demonstrate the cancellation
of the divergences for Fig.2 without having
to pass a $\gamma_5$ through a $\gamma_{\mu}$.
This is curiously different from the situation
in Fig.1.}

Please be aware that I will calculate the graphs in Figs.2-6
with the neglect of mixing between the heavy and light sectors.
I will comment in Sec.A.7 of this Appendix on the additional
contributions due to mixing.

The three graphs in Fig.2 combine to yield
\begin{eqnarray}
-i(\Sigma_{\tau}-\Sigma_b)|_H & = &
\int\frac{d^4k}{(2\pi)^4} \Bigg[
\frac{(a_H \sin\theta P_R + b_H P_L)(\hat{p}+\hat{k}
+ m_s)(a_H \sin\theta P_L + b_H P_R)}
{[(k + p)^2 - m_s^2]}\nonumber\\
& + & \frac{a_H^2 \cos^2\theta P_R(\hat{p}+\hat{k})P_L}
{(k + p)^2}\nonumber\\
& - &
\frac{(a_H P_R + b_H P_L)(\hat{p}+\hat{k} + m_t)
(a_H P_L + b_H P_R)}{[(k + p)^2 - m_t^2]} \Bigg] \times
\nonumber\\
& \times & \left( \frac{1}{k^2-m_H^2} -
\frac{1}{k^2 - \Lambda^2} \right).
\label{b-tauH}
\end{eqnarray}
Each of the three terms in this expression
are finite because of the Pauli-Villars regularisation.

Inspection of this equation reveals that the
potentially divergent part has an integrand
proportional to ${\rm div}_H$ where
\begin{eqnarray}
{\rm div}_H & = &
[a_H^2 \sin^2\theta (\hat{p}+\hat{k})P_L + b_H^2 P_R
+ m_s \sin\theta a_H b_H]\nonumber\\
& + &
[a_H^2 \cos^2\theta (\hat{p}+\hat{k})P_L]\nonumber\\
& - & [a_H^2 (\hat{p}+\hat{k}) P_L +
b_H^2 P_R + m_t a_H b_H].
\end{eqnarray}
Dirac algebra has been used to simplify this expression,
and the three terms in square brackets above
correspond to the three integrals in Eq.~(\ref{b-tauH}).
Using $m_s \sin\theta = m_t$ we see that ${\rm div}_H = 0$.

Taking $\Lambda \to \infty$ now that the divergences
have disappeared, and isolating the $\gamma_5$
part, I find that
\begin{eqnarray}
-i(\Sigma_{\tau}-\Sigma_b)|_H & = &
M^2 \int\frac{d^4k}{(2\pi)^4}
\frac{\frac{1}{2}b_H^2\hat{k} + m_t a_H b_H}
{[(k-p)^2 - m_H^2](k^2 - m_s^2)(k^2 - m_t^2)}
\nonumber\\
& + & \frac{1}{2} m_t^2 M^2 a_H^2
\int\frac{d^4k}{(2\pi)^4}
\frac{\hat{k}}{[(k-p)^2 - m_H^2]k^2(k^2 - m_s^2)(k^2 - m_t^2)}
\nonumber\\
& + & (\gamma_5\ {\rm part})
\label{SigmaH}
\end{eqnarray}
Integration variables have also been changed in this expression.

The integrals required above are the same as $I_3$, $\hat{I}_3$ and
$I_4$ introduced in the Sec.A.1 but with $\hat{p} \to -\hat{p}$. They
approximately evaluate to
\begin{eqnarray}
I_3 & \simeq & \frac{i}{16\pi^2}\frac{1}{m^2_H -
m_s^2}\ln\left(\frac{m_s^2}{m_H^2}\right),\\
\hat{I}_3 & \simeq & \frac{i}{32\pi^2}\frac{\hat{p}}{m_s^2-m_H^2}\left[
1 + \frac{m_s^2}{m^2_H - m_s^2}
\ln\left(\frac{m_s^2}{m_H^2}\right)\right],\\
\hat{I}_4 & \simeq & \frac{1}{m^2_s}\hat{I}_3,
\end{eqnarray}
under the condition that $m_H^2 \sim m_s^2 \gg m_t^2 \gg p^2 = m^2$.

The contributions to Eqn.~(\ref{SigmaH}) involving $\hat{I}_3$ and
$\hat{I}_4$ will generically be much smaller than that involving $I_3$.
The $\hat{k}$ in the integrand produces a $\hat{p}$ after integration
which in turn becomes an $m$ after the mass shift part is isolated. This
overall factor of $m$ is not cancelled off, as is evident from the
integral evaluations above, so this suppresses the
$\hat{I}$ terms relative to the $M^2 m_t a_H b_H I_3$ term. It is possible
to cancel the generically dominant term if $m_t\sin 2\omega \simeq m$.

Assuming this accidental cancellation does not occur, I find that
\begin{equation}
m_{\tau}-m_b|_{H} \simeq - \frac{1}{16\pi^2}
\frac{m_s^2 - m_t^2}{m_H^2 - m_s^2}
\frac{m_t(m_t - m\sin2\omega)(m_t\sin2\omega - m)}
{(u_1^2 + u_2^2)\cos^22\omega}
\ln\left(\frac{m_s^2}{m^2_H}\right).
\end{equation}

\vskip 3 mm

\centerline{\large \bf A.3 Graphs in Figure 3}

\vskip 2 mm

Using Pauli-Villars regularisation and working in
Feynman gauge, the graphs of
Fig.3 yield
\begin{eqnarray}
-i(\Sigma_{\tau} - \Sigma_b) & = &
\int\frac{d^4k}{(2\pi)^4}
\frac{m^2_{W_L} - \Lambda^2}{(k^2 - m^2_{W_L})(k^2 - \Lambda^2)}
\Bigg[ -
\frac{(\hat{k} + \hat{p})(a_g^2 P_R + b_g^2 P_L ) + a_g b_g m_t}
{(k + p)^2 - m_t^2} \nonumber\\
& + & \frac{(\hat{k} + \hat{p}) a_g^2 \cos^2\theta P_R}{(k+p)^2}
\nonumber\\
& + &
\frac{(\hat{k} + \hat{p})(a_g^2 \sin^2\theta P_R + b_g^2 P_L)
+ a_g b_g m_t}{(k+p)^2 - m_s^2} \Bigg],
\end{eqnarray}
where the three terms above correspond to the three graphs.
Dirac algebra simplification and $m_t = m_s\sin\theta$
have been used here.

The potentially divergent piece has an integrand proportional
to ${\rm div}_g$ where
\begin{eqnarray}
{\rm div}_g & = & [-(\hat{k} + \hat{p})(a_g^2 P_R + b_g^2 P_L)
- a_g b_g m_t] + [(\hat{k} + \hat{p})a_g^2 \cos^2\theta P_R]
\nonumber\\
& + & [(\hat{k} + \hat{p})
(a_g^2 \sin^2\theta P_R + b_g^2 P_L) + a_g b_g m_t].
\end{eqnarray}
The three terms in square brackets correspond to the
three graphs. Note that the divergences cancel.

Taking the cut-off to infinity, discarding the $\gamma_5$
term and changing integration variables reveals that
\begin{eqnarray}
-i(\Sigma_{\tau} - \Sigma_b) & = &
M^2\int\frac{d^4k}{(2\pi)^4}
\frac{\frac{1}{2} b_g^2 \hat{k} + a_g b_g m_t}
{[(k-p)^2 - m^2_{W_L}](k^2 - m_t^2)(k^2 - m_s^2)}
\nonumber\\
& + & \frac{1}{2} a_g^2 m_t^2 M^2
\int\frac{d^4k}{(2\pi)^4} \frac{1}
{[(k-p)^2 - m^2_{W_L}]k^2(k^2 - m_t^2)(k^2 - m_s^2)}
\nonumber\\
& + & (\gamma_5\ {\rm term}).
\end{eqnarray}
{}From the experience gained with the explicit evaluation of Figs.1 and 2
the qualitative behaviour of this expression can now be ascertained
without explicit computation.

In the limit $M^2 \to \infty$, the first term above
gives a large logarithm while the second does not. The first term thus
contributes to renormalisation
group running (plus residual threshold effects)
while the second term contains threshold effects only. By contrast with
Figs.1 and 2, however, the threshold effects will involve the mass
ratios of $W_L$ and $t$ which are relatively light particles.

None of these threshold terms are enhanced by $m_t$, however.
The potential $m_t^3$ term disappears
because of the chiral structure of the graphs.
To obtain such a term, a
$m_t b^2_g$ piece in the integrand would be needed. There is no such
term because it is proportional to $P_R P_L = 0$.
The potentially enormous $m_s m_t^2$ term is zero for
the same reason. I conclude therefore, that the low mass scale
threshold corrections from Fig.3 are numerically small compared to the
$m_t$ enhanced effects from Fig.2.

\vskip 3 mm

\centerline{\large \bf A.4 Graphs in Figure 4}

\vskip 2 mm

The three graphs in Fig.4 imply that
\begin{eqnarray}
-i(\Sigma_{\tau} - \Sigma_b)|_{W_L} & = &
\frac{g_L^2}{2} \int\frac{d^4k}{(2\pi)^4}
\frac{m^2_{W_L} - \Lambda^2}
{(k^2 - \Lambda^2)(k^2 - m^2_{W_L})} \Bigg[
\frac{\cos^2\theta
\gamma^{\mu}(\hat{k} + \hat{p})\gamma_{\mu}P_L}
{(k+p)^2}\nonumber\\
& + & \frac{\sin^2\theta
\gamma^{\mu}(\hat{k} + \hat{p})\gamma_{\mu}P_L}
{(k+p)^2 - m_s^2}
- \frac{\gamma^{\mu}(\hat{k} + \hat{p})\gamma_{\mu}P_L}
{(k+p)^2 - m_t^2}\Bigg]
\end{eqnarray}
where again Pauli-Villars regularisation has been used,
followed by Dirac algebra simplification. The three terms
above correspond to the three graphs in Fig.4.

It is easy to see by inspection that the potential
divergence cancels, giving that
\begin{equation}
-i(\Sigma_{\tau} - \Sigma_b)|_{W_L} =
\frac{g_L^2}{2} m_t^2 M^2 \int\frac{d^4k}{(2\pi)^4}
\frac{\gamma^{\mu}\hat{k}\gamma_{\mu} P_L}
{[(k-p)^2 - m^2_{W_L}]k^2 (k^2 - m_s^2)(k^2 - m_t^2)}.
\end{equation}
The cut-off has been taken to infinity and
integration variables changed to obtain this expression.
As $M^2 \to \infty$, this contribution remains finite.
Therefore it does not generate a large logarithm; it is purely a
(light mass scale) threshold effect.
The physical reason for this is that the divergence
cancellation cannot fail when the $\nu_R$ state is
removed from the physical spectrum by taking
$M^2 \to \infty$. The left-sector $W$ bosons couple
to $\nu_L$, so the absence of $\nu_R$ does not affect
the cancellation of divergences. There is also no enhancement
due to $m_t$, because the $m_t$ term in the numerator
disappears through $P_L P_R = 0$ and because the vertices
are not proportional to $m_t$.

\vskip 3 mm

\centerline{\large \bf A.5 Graphs in Figure 5}

\vskip 2 mm

The two graphs involving the $W_R$ boson lead to
\begin{eqnarray}
-i(\Sigma_{\tau} - \Sigma_b)|_{W_R} & = &
\frac{g_R^2}{2} \int\frac{d^4k}{(2\pi)^4}
\frac{m^2_{W_R} - \Lambda^2}{(k^2 - \Lambda^2)(k^2 - m^2_{W_R})}
\times \nonumber\\
& \times & \Bigg[
-\frac{\gamma^{\mu}(\hat{p} + \hat{k})\gamma_{\mu} P_R}
{(k + p)^2 - m_s^2}
+ \frac{\gamma^{\mu}(\hat{p} + \hat{k})\gamma_{\mu} P_R}
{(k + p)^2 - m_t^2} \Bigg]
\end{eqnarray}
where, again, Pauli-Villars regularisation and Dirac algebra
simplification have been used.

It is obvious that the potential divergence cancels between
the two graphs. Therefore it is clear that
\begin{equation}
-i(\Sigma_{\tau} - \Sigma_b)|_{W_R} =
-\frac{g_R^2}{2} M^2 \int\frac{d^4k}{(2\pi)^4}
\frac{\gamma^{\mu}\hat{k}\gamma_{\mu}P_R}
{[(k-p)^2 - m^2_{W_R}](k^2 - m_t^2)(k^2 - m_s^2)}.
\end{equation}
The cut-off has been taken to infinity and a change of
integration variables has been performed.

As the Pati-Salam breaking scale is taken to infinity, both
$M$ and $m_{W_R}$ go to infinity. In this limit then,
\begin{equation}
-i(\Sigma_{\tau} - \Sigma_b)|_{W_R} \to
- \frac{g_R^2}{2}\frac{1}{m^2_{W_R}}
\int\frac{d^4k}{(2\pi)^4}
\frac{\gamma^{\mu} \hat{k} \gamma_{\mu}}{k^2 - m_t^2}
\end{equation}
which integrates to zero because the integrand tends
to an odd function of $k$. Therefore no large logarithms are
generated by separating the two symmetry breaking scales and the terms
that remain nonzero for large but finite high scale masses are small.

\vskip 3 mm

\centerline{\large \bf A.6 Graphs in Figure 6}

\vskip 2 mm

I now turn to the diagrams involving the heavy Higgs bosons $\chi$.
I will again be able to demonstrate that the divergences cancel
without having to rearrange the Dirac matrices, so
I work in $n$-dimensions from the start.
The contribution of the unphysical Higgs boson $\chi^-$ is
\begin{equation}
-i\Sigma_{\tau}|_{\chi^-}  =
n^2 \int\frac{d^nk}{(2\pi)^n} \frac{1}{k^2 - m^2_{W_R}} \Bigg[
\frac{\sin^2\theta P_L (\hat{p} + \hat{k})P_R}{(k+p)^2}
 + \frac{\cos^2\theta P_L(\hat{p} + \hat{k})P_R}
{(k+p)^2 - m_s^2} \Bigg]
\end{equation}
where the $n$-dimensional result $P_L P_R = 0$ has been used.

The coloured boson $\chi^d$ on the other hand has a contribution
given by
\begin{equation}
-i\Sigma_b|_{\chi^d} =
n^2 \int\frac{d^nk}{(2\pi)^n} \frac{1}{k^2 - m^2_{\chi^d}} \Bigg[
\frac{\sin^2\theta P_L (\hat{p} + \hat{k})P_R}{(k+p)^2}
 + \frac{\cos^2\theta P_L(\hat{p} + \hat{k})P_R}
{(k+p)^2 - m_s^2} \Bigg],
\end{equation}
where again $P_L P_R = 0$ has been used and nothing more.

It is clear that the divergences cancel when the
$b$ contribution is subtracted
from the $\tau$ contribution. Deleting the
$\gamma_5$ part I find that
\begin{eqnarray}
-i(\Sigma_{\tau} - \Sigma_b)|_{\chi} & = &
\frac{n^2}{2} (m^2_{W_R} - m^2_{\chi^d}) \Bigg[
\sin^2\theta  \int\frac{d^4k}{(2\pi)^4}
\frac{\hat{k} + \hat{p}}
{(k^2 - m^2_{W_R})(k^2 - m^2_{\chi^d})(k + p)^2}
\nonumber\\
& + & \cos^2\theta  \int\frac{d^4k}{(2\pi)^4}
\frac{\hat{k} + \hat{p}}
{(k^2 - m^2_{W_R})(k^2 - m^2_{\chi^d})[(k + p)^2 - m_s^2]}\Bigg].
\end{eqnarray}
It is clear by inspection that these graphs produce high mass scale
threshold corrections, and that they are not enhanced by $m_t$.
\vskip 3 mm

\centerline{\large \bf A.7 Graphs in Figure 7}

\vskip 2 mm

All of the graphs in Fig.7 arise from mixing between the bosons
of the heavy sector with those of the light sector. They are all
individually finite.
A general argument shows that they cannot contribute
unsuppressed large logarithmic terms because they are
proportional to mixing angles between the heavy and light sectors.

Consider, for instance, a general Yukawa interaction of the
form
\begin{equation}
{\cal L} = \lambda_1 \overline{F} f S_1 + \lambda_2
\overline{f} F S_2 + {\rm H.c.}
\end{equation}
If the scalar bosons $S_1$ and $S_2$ do not mix, then they each
contribute separately to fermion self-energies via the
individually divergent diagrams I have been considering. However,
if they mix with a mixing angle $\zeta$, then
\begin{equation}
{\cal L} = \lambda_1 \overline{F} f (\cos\zeta S'_1 + \sin\zeta S'_2)
+ \lambda_2
\overline{f} F (-\sin\zeta S'_1 + \cos\zeta S_2) + {\rm H.c.}
\end{equation}
where the primed fields denote the new mass eigenstates. This gives
rise to a new contribution proportional
to the mixing parameters.\footnote{Note that when mixing is considered
the graphs I have already calculated which do not require mixing
to exist will be multiplied by $\cos^2\zeta \simeq 1$ factors.}
For instance, the self-energy of $f$ receives an additional
finite contribution given by
\begin{eqnarray}
-i\Sigma_f & = & -\lambda_1\lambda_2 \sin\zeta \cos\zeta
\int\frac{d^4k}{(2\pi)^4}
\Bigg(\frac{1}{k^2-m_1^2}-\frac{1}{k^2-m_2^2}\Bigg)
\frac{\hat{k}+\hat{p}+m_F}{(k+p)^2-m_F^2}
\nonumber\\
& = &  -\lambda_1\lambda_2 \sin\zeta \cos\zeta
(m_1^2 - m_2^2)\times\nonumber\\
& \times & \int\frac{d^4k}{(2\pi)^4}
\frac{\hat{k}+\hat{p}+m_F}{(k^2-m_1^2)(k^2-m_2^2)
[(k+p)^2-m_F^2]}
\end{eqnarray}
where $m_{1,2}$ is the mass of $S'_{1,2}$, and $m_F$ is the mass
of $F$. Suppose the heavy scalar to be $S'_2$. In the limit
that $m_2 \to \infty$,
\begin{equation}
-i\Sigma_f \to -\lambda_1\lambda_2 \sin\zeta \cos\zeta
\int\frac{d^4k}{(2\pi)^4}
\frac{\hat{k}+\hat{p}+m_F}{(k^2-m_1^2)[(k+p)^2-m_F^2]}.
\end{equation}
The integral above is logarithmically divergent and thus there
will be a large logarithm in the heavy mass $m_2$. However, the
self-energy is also proportional to $\sin\zeta \cos\zeta$, which
goes to zero as the heavy scale is taken to infinity. Generically,
mixing angles between heavy and light scalars go as
at most $m_{\rm light}/m_{\rm heavy}$ as the heavy mass goes to
infinity. Therefore the large logarithm above will always be
suppressed by $m_1/m_2$ and thus it will be ineffective.

Note that the statement that the mixing angle will generically
go as $m_{\rm light}/m_{\rm heavy}$ is not the same as the
statement that we always want one light eigenstate and one
heavy eigenstate. For instance, a ``democratic'' $2\times 2$
mass matrix
(which has each entry as 1) will yield one zero and one nonzero
eigenvalue but with a mixing angle of $\pi/4$. However, in this case
there is no clear separation of the unmixed fields into a heavy
and a light sector. One must make sure that the model does not
produce this type of situation. This means that
a scalar mass hierarchy must be put into the theory by hand
and then preserved
to all orders of perturbation theory (at least).
This is of course just a manifestation of the gauge hierarchy
problem for scalar bosons.

The argument above may be easily repeated for graphs dependent on
gauge boson mixing instead of scalar boson mixing.

\newpage

\centerline{\Large \bf Figure Captions}

\vskip 1 cm

\noindent
{\bf Figure 1:} Feynman graphs contributing to
$m_{\tau}-m_b$ which involve the photon $\gamma$, the
$Z$, $Z'$ and $X$ bosons and the gluons $G$. The logarithmic
divergences of the individual self energies cancel in
$m_{\tau}-m_b$ between these graphs. The external fermion line
is either $\tau$ or $b$ for the $\gamma$, $Z$, $Z'$ and $X$
graphs, while the external fermion for the gluon graph is $b$ only.
The internal fermion for the $\gamma$, $Z$ and $Z'$ graphs is the
same as the external fermion. For the $X$ graph, the internal fermion
is a $\tau(b)$ if the external fermion is a $b(\tau)$. The internal
fermion for the gluon graph is a $b$. In section A.1 of the Appendix,
I calculate $m_{\tau}-m_b$ under the approximation that $m_Z = 0$.
This allows a change from the $(\gamma, Z)$ basis to the
$(W^0, B)$ basis. The $W^0$ boson graph does not contribute to
$m_{\tau}-m_b$ because $W^0$ couples universally to $b$ and $\tau$.
In the text I therefore actually calculate the four diagrams
involving a massless $B$ boson, the $Z'$ and $X$ bosons, and the
gluons.

\vskip 5 mm

\noindent
{\bf Figure 2:} Feynman graphs involving the physical charged
Higgs boson $H^-$. The individual divergences cancel in
$m_{\tau}-m_b$ between these graphs.

\vskip 5 mm

\noindent
{\bf Figure 3:} Feynman graphs involving the unphysical charged
Goldstone boson $g^-$. The individual divergences cancel in
$m_{\tau}-m_b$ between these graphs.

\vskip 5 mm

\noindent
{\bf Figure 4:} Feynman graphs involving the left-sector gauge boson
$W_L^-$. The individual divergences cancel in
$m_{\tau}-m_b$ between these graphs.

\vskip 5 mm

\noindent
{\bf Figure 5:} Feynman graphs involving the right-sector
gauge boson $W_R^-$. The divergences cancel in
$m_{\tau}-m_b$ between these two graphs.

\vskip 5 mm

\noindent
{\bf Figure 6:} Feynman graphs involving components of $\chi$.
The divergences cancel in $m_{\tau}-m_b$ between these two graphs.

\vskip 5 mm

\noindent
{\bf Figure 7:} The first two Feynman graphs contribute to
$m_{\tau}-m_b$ when $W_L^- - W_R^-$ mixing is switched on.
The third graph contributes when $Z-Z'$ mixing is included.
The fourth
graph denotes the fact that the Goldstone bosons eaten by $W_R^-$
and $W_L^-$ are actually linear combinations of $\chi^-$ and $g^-$.

\newpage
\centerline{\epsfbox{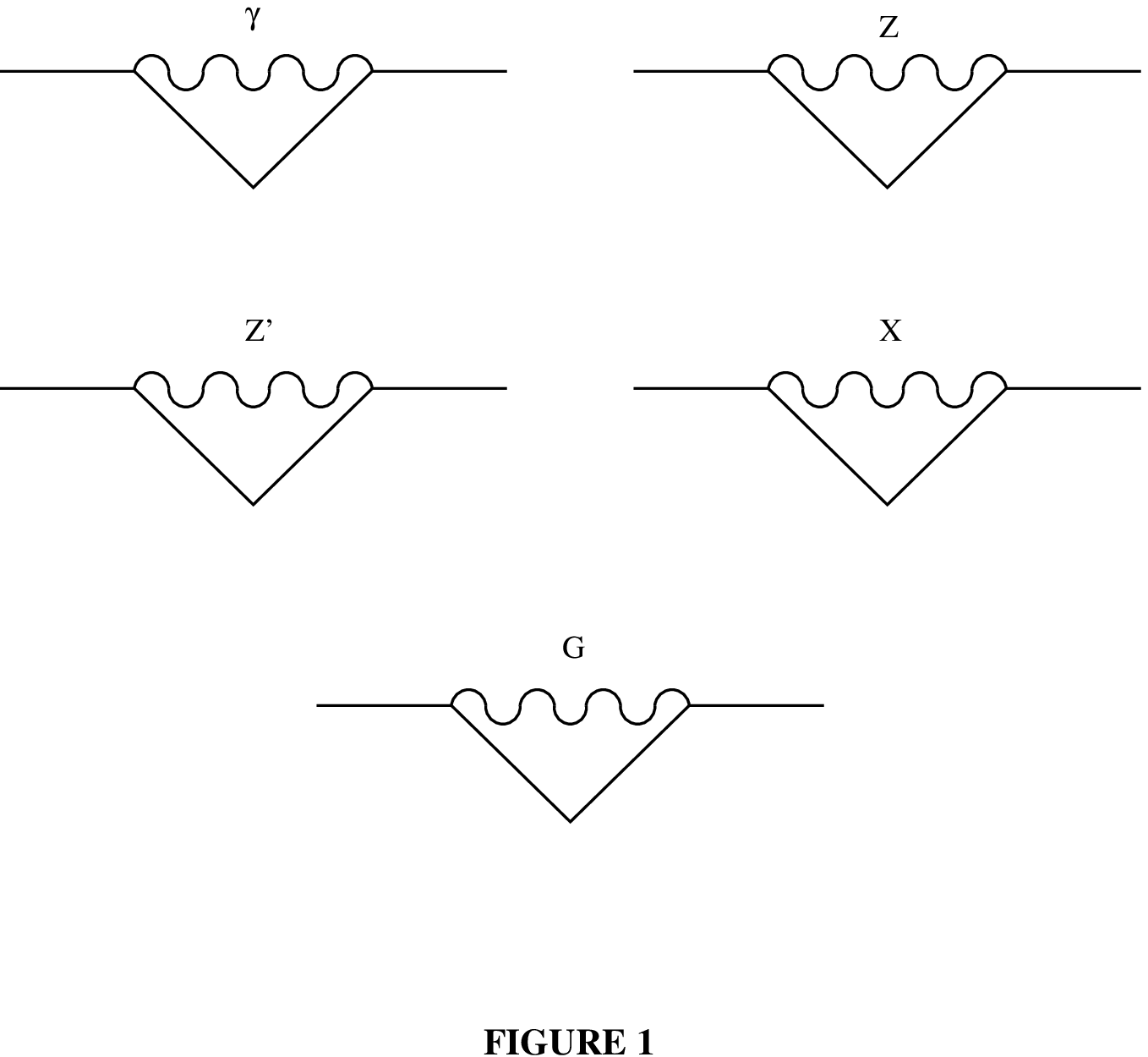}}
\centerline{\epsfbox{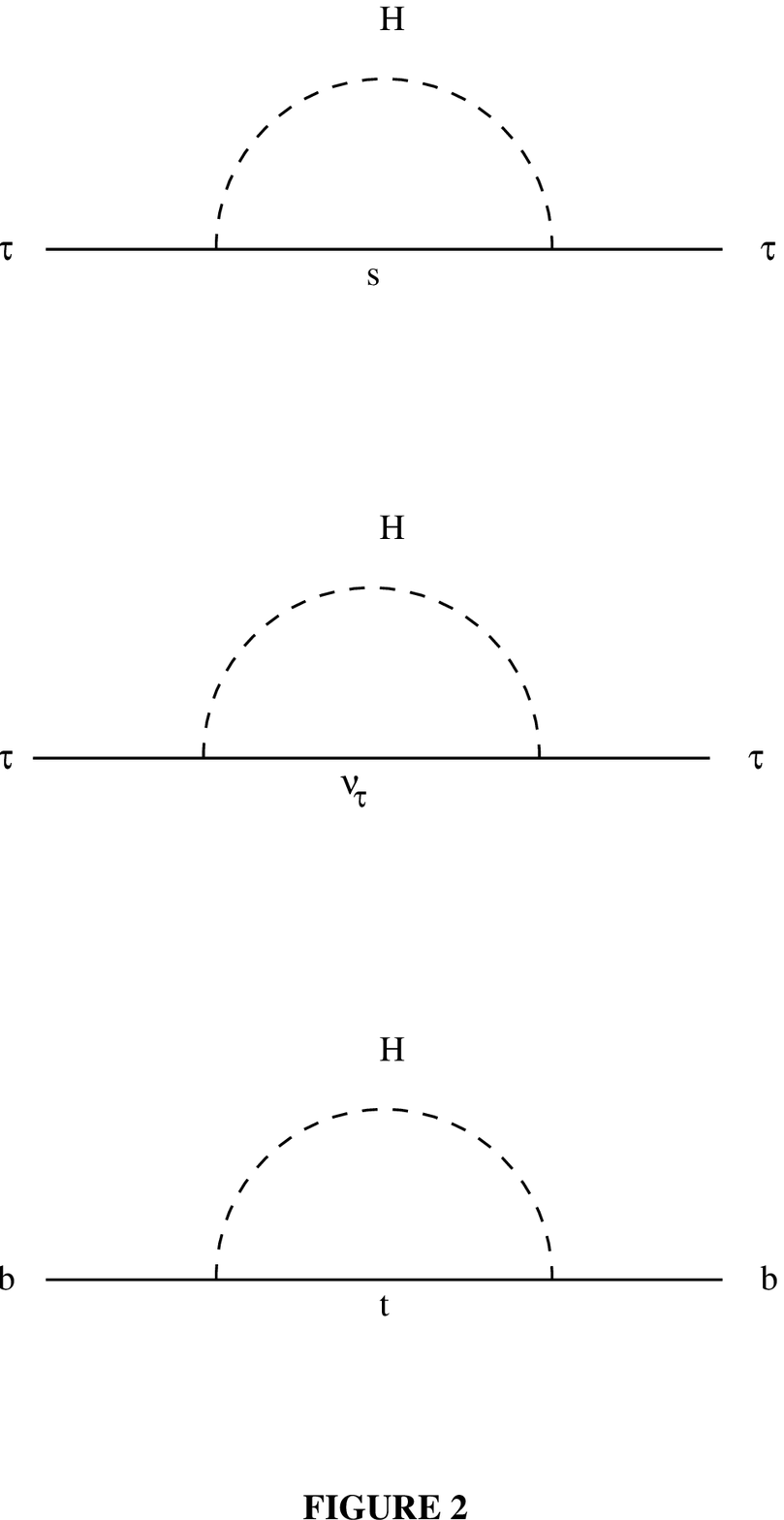}}
\centerline{\epsfbox{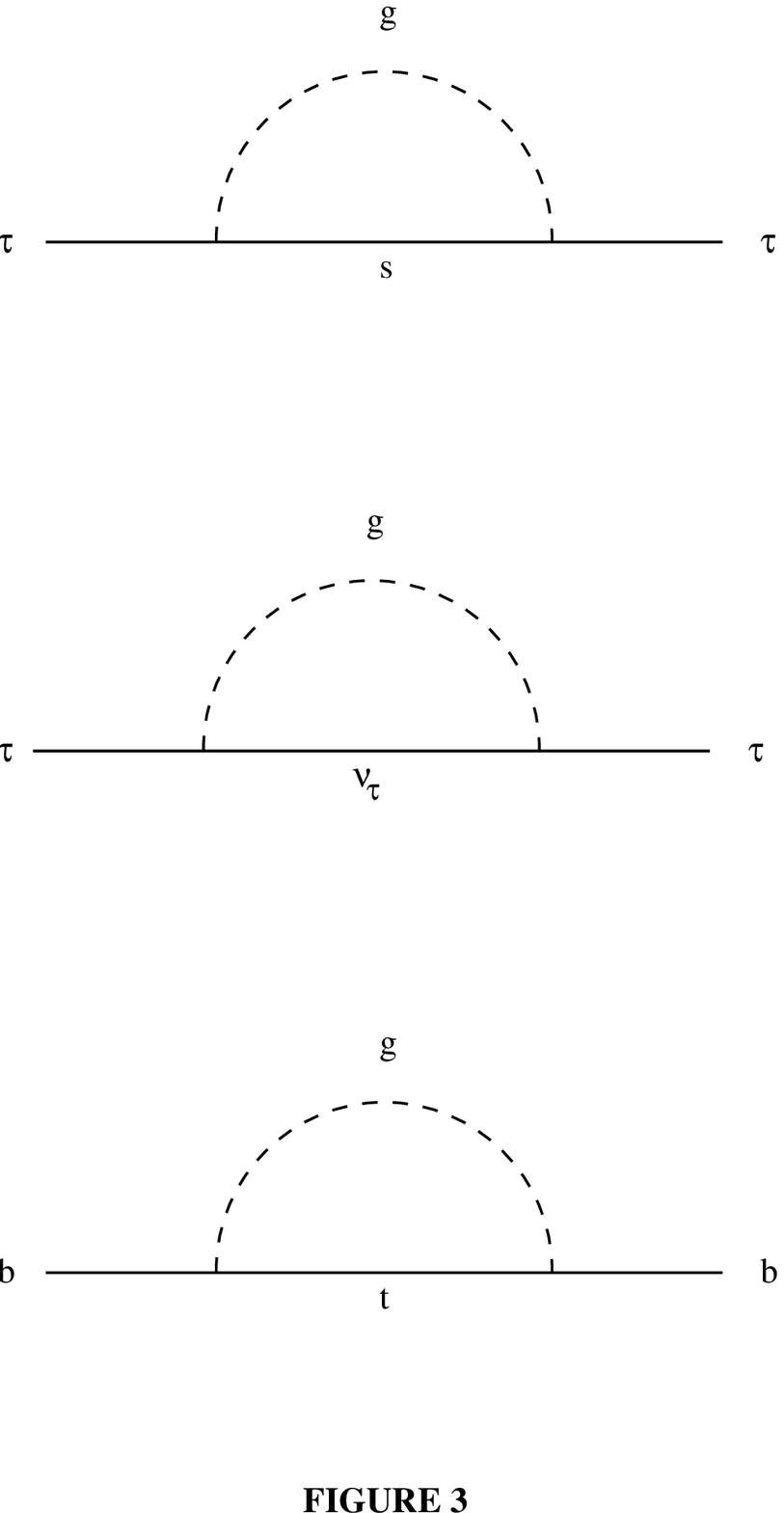}}
\centerline{\epsfbox{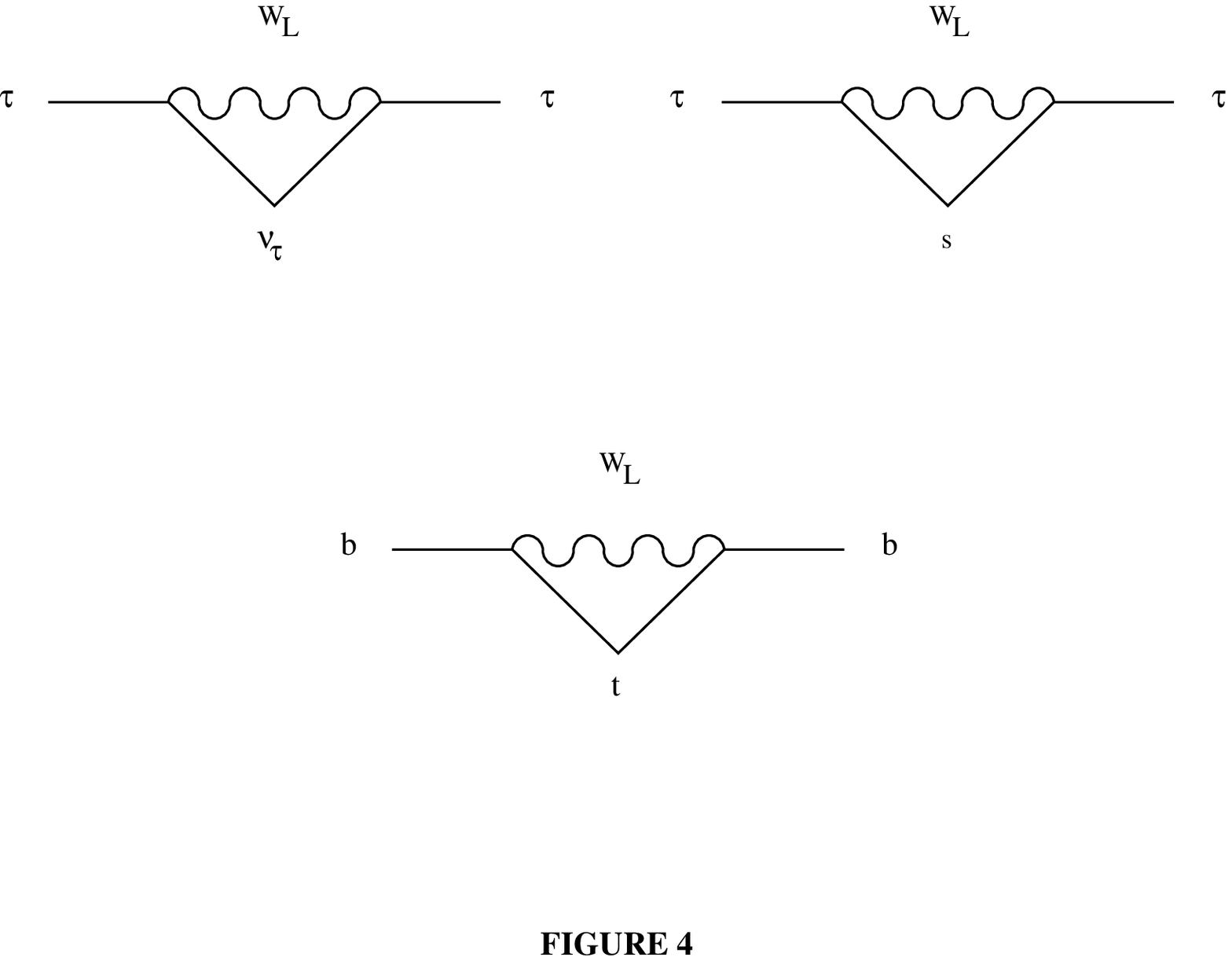}}
\centerline{\epsfbox{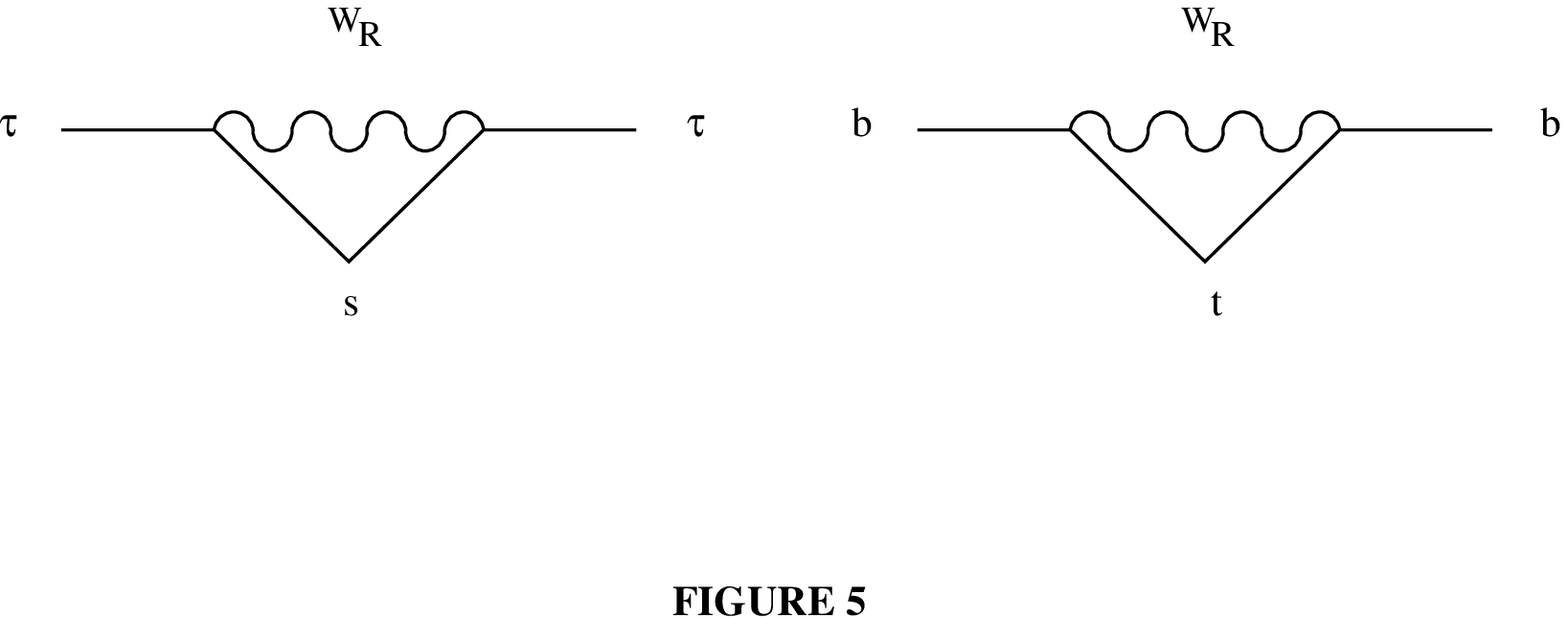}}
\vskip 3 cm
\centerline{\epsfbox{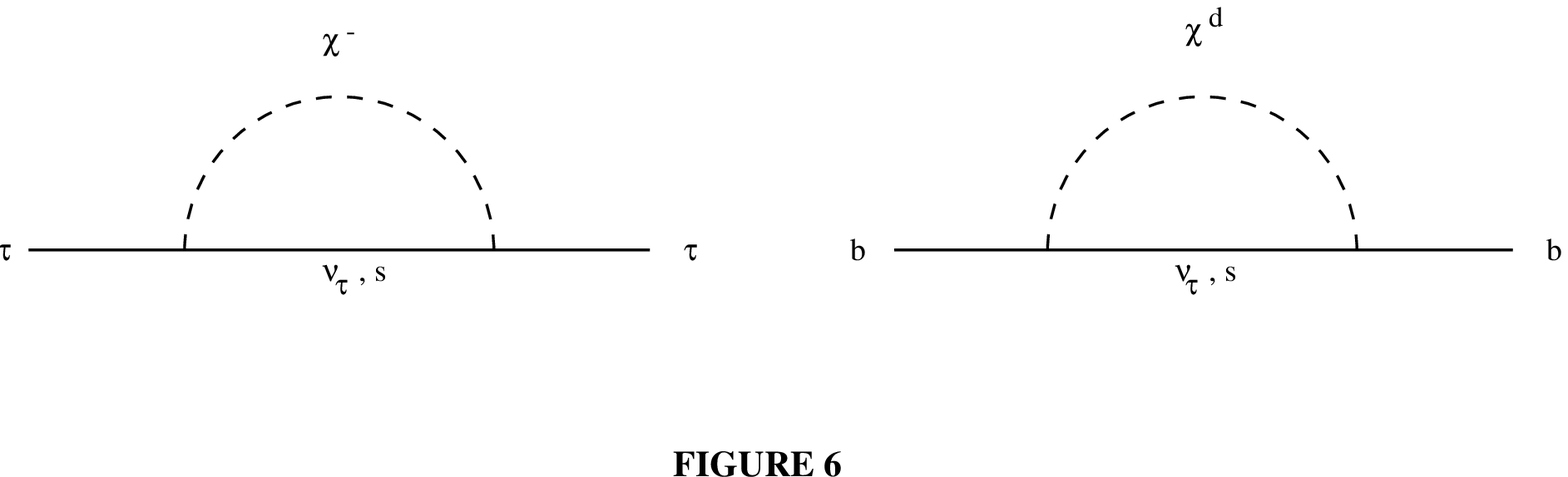}}
\centerline{\epsfbox{fig7.ps}}

\newpage
\begin{thebibliography}{99}

\bibitem{weinberg} This group theoretic idea was noted in a
slightly different context long ago. See S. Weinberg,
Phys.\ Rev.\ Lett.\ {\bf 29}, 388 (1972).

\bibitem{ps} J. C. Pati and A. Salam, Phys.\ Rev.\ {\bf D8}, 1240
(1973); {\it ibid.} {\bf D10}, 275 (1974).

\bibitem{ql} R. Foot and H. Lew, Phys.\ Rev.\ {\bf D41}, 3052 (1990);
R. Foot, H. Lew and R. R. Volkas, {\it ibid.} {\bf D44}, 1531 (1991);
R. R. Volkas, {\it ibid.} {\bf D50}, 4625 (1994).

\bibitem{willenbrock} G. Valencia and S. Willenbrock, Phys.\ Rev.\
{\bf D50}, 6845 (1994).

\bibitem{rnm} See, for instance, R. N. Mohapatra,
{\it Unification and Supersymmetry}, (Springer-Verlag, New York, 1986).

\bibitem{Parida} D. Chang, R. N. Mohapatra and M. K. Parida,
Phys.\ Rev.\ Lett.\ {\bf 52}, 1072 (1984).

\bibitem{seesaw} T. Yanagida, in {\it Proceedings of the Workshop on
Unified Theory and Baryon Number of the Universe}, edited by
O. Sawada and A. Sugamoto (KEK, Tsukuba, Japan, 1979);
M. Gell-Mann, P. Ramond and R. Slansky, in {\it Supergravity},
edited by P. Van Nieuwenhuizen and D. Freedman (North-Holland,
Amsterdam, 1980); R. N. Mohapatra and G. Senjanovic, Phys.\ Rev.\
Lett.\ {\bf 44}, 912 (1980).

\bibitem{kt} See, for instance, p. 239 of
E. W. Kolb and M. S. Turner, {\it The Early Universe},
(Addison-Wesley, Redwood City, 1990).

\bibitem{ramond} H. Arason et al., Phys.\ Rev.\ {\bf D46}, 3945 (1992).

\bibitem{babu} See, for instance, K. S. Babu, K. Fujikawa and
A. Yamada, Phys.\ Lett.\ {\bf B333}, 196 (1994).

\bibitem{pati} See D. Chang, R. N. Mohapatra, P. B. Pal and J. C. Pati,
Phys.\ Rev.\ Lett.\ {\bf 55}, 2756 (1985)
for a discussion of such a possibility when
the top quark was thought to be much lighter.

\bibitem{ma} E. Ma, Phys.\ Rev.\ {\bf D36}, 271 (1987);
K. S. Babu, X.-G. He and E. Ma, {\it ibid.} {\bf D36}, 878 (1987).

\bibitem{gj} H. Georgi and C. Jarlskog, Phys.\ Lett.\ {\bf B86}, 297
(1979).

\bibitem{worah} M. Worah, Enrico Fermi Institute report EFI-95-04
(unpublished), hep-ph/9502222.
\end{thebibliography}
\end{document}